\documentclass[%
 reprint,
 amsmath,amssymb,
 aps,
]{revtex4-1}

\usepackage{graphicx}
\usepackage{dcolumn}
\usepackage{bm}

\begin{document}


\title{First-passage process in degree space for the time-dependent Erd\H{o}s-R\'enyi and Watts-Strogatz models}

\author{F. Ampuero and M. O. Hase}
 \affiliation{Escola de Artes, Ci\^encias e Humanidades, Universidade de S\~ao Paulo, Av. Arlindo B\'ettio 1000, 03828-000 S\~ao Paulo, Brazil }

\date{\today}

\begin{abstract}
In this work, we investigate the temporal evolution of the degree of a given vertex in a network by mapping the dynamics into a random walk problem in degree space. We analyze when the degree approximates a pre-established value through a parallel with the first-passage problem of random walks. The method is illustrated on the time-dependent versions of the Erd\H{o}s-R\'enyi and Watts-Strogatz models, which originally were formulated as static networks. We have succeeded in obtaining an analytic form for the first and the second moments of the first-passage time and showing how they depend on the size of the network. The dominant contribution for large networks with $N$ vertices indicates that these quantities scale on the ratio $N/p$, where $p$ is the linking probability.
\end{abstract}

\maketitle



\section{\label{introdution}Introduction}

The study of the properties of networks led to the development of many mathematical, statistical and computational tools that can be used to analyze, model, and understand how systems behave in many areas of knowledge such as physics, biology, ecology, and social sciences, to name some of them. Modeling complex systems by networks \cite{AB02, DM03, N10} is a natural strategy to investigate a system from a very basic structure composed of agents and interactions among them, represented, respectively, by vertices (or nodes) and links.

In this work, we are mainly interested in the time evolution of the degree of a given node. Concretely, a vertex can gain and/or lose connections during its dynamics, and we investigate when it achieves a pre-established degree for the first time. This is a particularly relevant issue when agents can not afford an indefinite number of connections and some indication of approaching the maximum capacity of the node \cite{ASBS00, HCG16} is desired. As an instance, it is known that airports (where the links can be assigned to the routes) have constraints that prevent growth without careful planning \cite{UTGR03}.

We map the dynamics of increasing/decreasing degrees into a random walk process, as was introduced in \cite{CGCH17}, and see if and how long it takes for a vertex with degree $k_{0}$ to achieve degree $k$ for the first time. This is a one-dimensional random walk in degree space, where the rules of gaining/losings degree are governed by the dynamics of the network. The random walk is a classical problem \cite{R80, P05} where a particle moves in random directions and one typically inquires about its statistical properties after a long time. Starting from an origin, one possible question that can be formulated concerns the probability of returning to the starting point, and the first-passage process refers to its return for the first time \cite{R07}. First-passage processes are seen in many applications, and examples are present in fluorescence quenching, integrate-and-fire neurons, and triggering of stock options, to cite some of them \cite{KRBN10, R07}. When the random walk is defined on a (hyper)cubic lattice \cite{P21}, and the particle can move in any direction with the same probability, it is known that the mean first-passage time scales as $L^{d}$ \cite{MW65}, where $L$ is the linear size of the $d$-dimensional lattice with periodic boundary conditions; furthermore, in the limit of infinite lattice, this random walk is known to be recurrent (\textit{i.e.}, it returns to the origin with probability $1$) for one-dimensional chains and two-dimensional square lattices, while the process is transient (\textit{i.e}, there is a positive probability of not returning to the origin) for hypercubic lattices with larger dimensions \cite{MW65, KRBN10}.

We investigate how the mean first-passage time of a vertex to reach a pre-established degree scales with the size of the network and other relevant parameters. Our study is based on an extended version of the Watts-Strogatz model \cite{WS98}, which was also introduced in \cite{CGCH17}, but we consider first a dynamical version of the Erd\H os-R\'enyi model \cite{ER59} to illustrate and outline the main steps of analysis. The choice of these two models is justified by a simplification that arises from a property shared between them, which is the time-translational invariance. Systems that do not have this property, like the random recursive tree \cite{NR70} or Barab\'asi-Albert network \cite{BA99}, indicate the need for a different approach, and will be examined elsewhere.

This paper is organized as follows. We define the dynamical version of the Erd\H{o}s-R\'enyi and Watts-Strogatz models in Section II and the general formalism to investigate the moments of the first-passage time is presented in Section III. The results for both models are shown in Section IV and some final comments are given in the last section.


\section{\label{models}Models}

Two models are introduced to test our ideas in this work. Both of them are already well-known in the literature \cite{ER59, WS98}, but were initially defined as static networks.

The first one, which is a minimal model, is the dynamical Erd\H os-R\'enyi network: the dynamics is just a simple addition of edges per time unit, and we monitor the increase of degrees only. The second one, the dynamical Watts-Strogatz model, is the simplest network that contains the process where a vertex can gain and/or lose connections randomly. The usual first-passage process (which is concerned with the return to the starting point) in the latter model corresponds to the so-called Motzkin paths \cite{OvdJ15}.


\subsection{Time-dependent Erd\H{o}s-R\'enyi model}

In the dynamical version of the Erd\H{o}s-R\'enyi model, consider a network with $N$ vertices. At each unitary time step, two vertices are randomly chosen and connected with probability $p$; this includes the possibility of (a) having a self-loop (\textit{i.e}, an edge that connects a vertex to itself) and (b) having more than one connection between the same pair of vertices. Since there is no preferential attachment, the probability of any vertex being chosen is $1/N$.

Defining $p_{s}(k,t)$ as the probability that a vertex $s$ has degree $k$ at time $t$, the dynamics can be represented by the recurrence relation
\begin{align}
\nonumber p_{s}(k,t+1) &= \omega_{\text{\tiny ER}}(k|k-2)p_{s}(k-2,t)+\\
\nonumber & + \omega_{\text{\tiny ER}}(k|k-1)p_{s}(k-1,t) + \\
 & + \omega_{\text{\tiny ER}}(k|k)p_{s}(k,t).
\label{rr_ER}
\end{align}
The term $\omega_{\text{\tiny ER}}(k|m)$ is the time-independent transition rate of changing the degree of a vertex from $m$ to $k$; in this time-discrete case with unitary time step, the transition rate coincides numerically to the conditional probability. The right-hand-side of the dynamics \eqref{rr_ER} contemplates three cases:
\renewcommand\labelenumi{(\roman{enumi})}
\begin{enumerate}
\item The degree of vertex $s$ changes from $k-2$ (at time $t$) to $k$ (at time $t+1$). An edge is introduced, with probability $p$ (there should be no confusion with $p_{s}$), the vertex $s$ is chosen twice and is connected to itself; this leads to
\begin{align}
\omega_{\text{\tiny ER}}(k|k-2) = \frac{p}{N^{2}};
\label{w(k-2)_ER}
\end{align}

\item The degree of vertex $s$ changes from $k-1$ (at time $t$) to $k$ (at time $t+1$). An edge is introduced, with probability $p$, and links to two different vertices: the vertex $s$ is just one of them. This situation is described by
\begin{align}
\omega_{\text{\tiny ER}}(k|k-1) = \frac{2p}{N}\left(1 - \frac{1}{N}\right);
\label{w(k-1)_ER}
\end{align}

\item The vertex $s$ already has degree $k$, and one should consider the probability of not changing its degree, \textit{i.e.}, the link is not introduced (with probability $1-p$) or, when the edge joins the network (with probability $p$), it connects two vertices other than $s$ with probability $\left(1-1/N\right)^{2}$. In this case, one has
\begin{align}
\nonumber \omega_{\text{\tiny ER}}(k|k) &= \left(1- p\right) + p\left(1-\frac{1}{N}\right)^{2} \\
&=1 - \frac{2p}{N} + \frac{p}{N^{2}}.
\label{w(k)_ER}
\end{align}
\end{enumerate}



\subsection{\label{sec:level3}Time-dependent Watts-Strogatz model}

In this version of the Watts-Strogatz model, the network has a fixed number $N$ of vertices and
\begin{align}
M := cN
\label{M}
\end{align}
degrees, where $c$ is the mean degree of the network (therefore, the entire graph has $cN/2$ edges). At each time step, an edge end is chosen at random with uniform probability $1/M$ and reconnected with probability $p$ (and no action takes place with probability $1-p$). This scheme does not forbid self-loops.

Defining $p_{s}(k,t)$ as the probability that a vertex $s$ has degree $k$ at time $t$ as before, the dynamics can be represented by
\begin{align}
\nonumber p_{s}(k,t+1) &= \omega_{\text{\tiny WS}} (k|k-1)p_{s}(k-1,t) + \\
\nonumber & + \omega_{\text{\tiny WS}} (k|k+1)p_{s}(k+1,t) + \\
& + \omega_{\text{\tiny WS}} (k|k)p_{s}(k,t),
\label{rr_WS}
\end{align}
where $\omega_{\text{\tiny WS}} (k|m)$ represents the time-independent transition rate of a vertex changing its degree from $m$ to $k$.

There are some different possible scenarios for a given vertex to change its degree from $m$ to $k$ in a single time step:
\renewcommand\labelenumi{(\roman{enumi})}
\begin{enumerate}
\item The degree of vertex $s$ changes from $k-1$ (at time $t$) to $k$ (at time $t+1$). An edge end not connected to $s$ is chosen with probability $1-\frac{k-1}{M}$, rewired with probability $p$ and connects to $s$ with probability $\frac{1}{N}$. This gives
\begin{align}
\omega(k | k-1) = \frac{p}{N}\left(1-\frac{k-1}{M}\right);
\label{w(k-1)_WS}
\end{align}

\item The degree of vertex $s$ changes from $k+1$ (at time $t$) to $k$ (at time $t+1$). An edge end connected to $s$ is chosen with probability $\frac{k+1}{M}$, rewired with probability $p$ and connects to a vertex other than $s$ with probability $1-\frac{1}{N}$, resulting in
\begin{align}
\omega(k | k+1) = \frac{k+1}{M} p \left(1-\frac{1}{N}\right);
\label{w(k+1)_WS}
\end{align}

\item The vertex $s$ has degree $k$ at time $t$ and neither gains or loses connections. This is represented by the sum of some disjoint cases: (a) there is no rewiring at all in the process with probability $1-p$, or (b) an edge end connected to $s$ is chosen with probability $\frac{k}{M}$, rewired with probability $p$ and connected again to $s$ with probability $\frac{1}{N}$; (c) an edge end not connected to $s$ is chosen with probability $1-\frac{k}{M}$ and rewired (with probability $p$) to connect to a vertex othen than $s$ with probability $1-1/N$. The sum of these probabilities results in
\begin{align}
\nonumber \omega(k|k) &= \left(1-p\right) + p\frac{k}{M}\frac{1}{N} + p\left(1-\frac{k}{M}\right)\left(1-\frac{1}{N}\right) \\
&= 1 - \frac{p}{N}\left(1+\frac{kN}{M} - \frac{2k}{M}\right).
\label{w(k)_WS}
\end{align}    
\end{enumerate}


\section{\label{rwds}Random walk in degree space}

Considering that vertices, in general, gain or lose connections, one can look at these changes in degree (of a specified vertex) as a one-dimensional random walk in degree space \cite{CGCH17}. Furthermore, the mean time required by a vertex to achieve a certain degree for the first time can be evaluated through a parallel with the first-passage problem of random walks \cite{R07, KRBN10}.

In both models presented in the previous section, there are two important symmetries. Firstly, the particular choice of a vertex $s$ is irrelevant, and this parameter has no role in our work - except for remembering that we are dealing with the time evolution of the degree of a given vertex.

The mean time $\langle t \rangle$ to achieve a certain degree $k$ for the first time (starting from $k_{0}$ at time $t_{0} = 0$) is given by
\begin{align}
\langle t \rangle = \sum_{t=0}^{\infty} tf_{s}(k,t|k_0,0),
\label{<t>}
\end{align}
where $f_{s}(k,t|k_{0},0)$ is the probability of vertex $s$ having degree $k$ for the first time at $t$, given that it had degree $k_{0}$ at time $t_{0}=0$. This probability can be obtained from the discrete time version of the first-passage process equation \cite{R07, KRBN10}, and can be cast as
\begin{align}
p_{s}(k,t|k_{0},0) = \sum_{t^{\prime}=0}^{t}f_{s}(k,t^\prime|k_{0},0) p_{s}(k,t|k,t^\prime), 
\label{bare_fp}
\end{align}
which describes the probability $p_{s}(k,t|k_{0},0)$ of the vertex $s$ having degree $k$ at time $t$ (not necessarily for the first time), given that it had degree $k_{0}$ at time $t_{0}=0$. This is a sum of all disjoint probabilities where the degree of the vertex achieves degree $k$ at time $t^{\prime}$ ($\leq t$) for the first time, and then reaches degree $k$ again at instant $t$. The initial condition $p_{s}(k,0|k_{0},0)=\delta_{k,k_{0}}$ is satisfied by assuming $f_{s}(k,0|k_{0},0)=\delta_{k,k_{0}}$ (an extra term in \eqref{bare_fp} associated to the initial condition is not required here as it is in the continuous-time version \cite{R07, KRBN10} of the equation).

The second important symmetry of our models can be seen from the transitions rates $\omega_{\text{\tiny ER}}$ and $\omega_{\text{\tiny WS}}$: they are invariant under time translation. As a consequence, $p_{s}(k,t|k^{\prime},t^{\prime})=p_{s}(k|k^{\prime};t-t^{\prime})$ and $f_{s}(k,t|k^{\prime},t^{\prime})=f_{s}(k|k^{\prime};t-t^{\prime})$ depend on the difference $t-t^{\prime}$ only. Therefore, equation \eqref{bare_fp} can be cast as
\begin{align}
p_{s}(k|k_{0};t) = \sum_{t^{\prime}=0}^{t}f_{s}(k|k_{0};t^{\prime}) p_{s}(k|k;t-t^\prime).
\label{fp}
\end{align}

As usual, the convolution product in \eqref{fp} suggests the introduction of the characteristic function
\begin{align}
p_{s}^{z}(k|k_0;z) = \sum_{t=0}^{\infty} z^{t}p_{s}(k|k_0;t)
\label{characteristic_t}
\end{align}
and a similar definition for the characteristic function of the function $f_{s}$. Then, it is immediate that
\begin{align}
f_{s}^{z}(k|k_0;z) = \frac{p_{s}^{z}(k|k_{0};z)}{p_{s}^{z}(k|k;z)},
\label{fz_pz}
\end{align}
and we can obtain $f_{s}^{z}$ from $p_{s}^{z}$. As stated before, this is a consequence of the time-translation invariance; models that do not have this symmetry (like the random recursive tree \cite{R80, P05} or Barab\'asi-Albert network \cite{BA99}) do not display the form \eqref{fp}.

We are mainly interested in \eqref{fz_pz} because it provides some quantities of interest. The first one is
\begin{align}
\mathcal{A} := \lim_{z\rightarrow 1}f_{s}^{z}(k|k_0;z) = \sum_{t=0}^{\infty} f_{s}(k|k_0;t),
\label{A}
\end{align}
which stands for the arriving probability of a vertex achieving degree $k$, starting from degree $k_{0}$, at some time, while
\begin{align}
\langle t^{n}\rangle = \lim_{z\rightarrow 1}\,\left(z\partial_z\right)^{n}\,f_{s}^{z}(k|k_0;z) = \sum_{t=0}^{\infty}t^{n}f_{s}(k|k_{0};t),
\label{<tn>}
\end{align}
where $\partial_{z}$ stands for the partial derivation in $z$ variable, shows that the quantity $f_{s}^{z}$ is also useful to evaluate any moment of the first-passage time. In this work, we are particularly interested in the first and second moments, $\langle t\rangle$ and $\langle t^{2}\rangle$, respectively; the latter is directly associated to the variance $\sigma^{2}=\langle t^{2}\rangle-\langle t\rangle^{2}$.

Hence, one can also expand \eqref{fz_pz} as
\begin{align}
f_{s}^{z}(k|k_{0};z) = \mathcal{A} + \langle t\rangle\left(z-1\right) + \left[\frac{\langle t^{2}\rangle-\langle t\rangle}{2}\right]\left(z-1\right)^{2} + \cdots
\label{exp_fz}
\end{align}
and obtain the desired quantites ($\mathcal{A}$, $\langle t\rangle$ and $\langle t^{2}\rangle$) through this representation.


\section{\label{results}Results}

In this section, we present the results for the first and second moments of the first-passage time for both models.

\subsection{\label{ER}Time-dependent Erd\H{o}s-R\'enyi model}

The discrete time evolution for the dynamical version of Erd\H{o}s-R\'enyi model, introduced in section \ref{models}, is given by \eqref{rr_ER}. Introducing the characteristic function
\begin{align}
p_{s}^{K}(K; t) = \sum_{k=0}^{\infty}K^{k}p_{s}(k|k_{0}; t)
\label{characteristic_K}
\end{align}
into \eqref{rr_ER} leads to
\begin{align}
p_{s}^{K}(K,t) = \left[ 1 - p + p\,\left(\frac{K}{N} + 1 - \frac{1}{N}\right)^{2} \right]^{t}\,K^{k_{0}},
\label{pK_ER}
\end{align}
where the initial condition $p_{s}(k|k_{0}; 0)=\delta_{k,k_{0}}$ or, equivalently, $p_{s}^{K}(K; 0)=K^{k_{0}}$ was adopted.
From \eqref{characteristic_K}, the probability $p_{s}(k|k_{0}; t)$ is the coefficient of the term $K^{k}$ in the series; therefore, expanding \eqref{pK_ER} and organizing the terms implies
\begin{align}
\nonumber p_{s}(k|k_{0}; t) &= \sum_{m = \left\lceil\frac{\Delta}{2}\right\rceil}^{t} {t \choose m}{2m\choose \Delta}\left(1-p\right)^{t-m}p^{m}\times\\
&\times\left(1-\frac{1}{N}\right)^{2m-\Delta}\frac{1}{N^{\Delta}}.
\label{p_ER}
\end{align}
From \eqref{p_ER}, the function $p_{s}$ depends on the difference $\Delta:=k-k_{0}$ only, and not on the initial and final degrees independently. This property is propagated to the quantities of interest in this work.

Using \eqref{p_ER}, the characteristic function (in time variable) of $p_{s}$ is
\begin{align}
\nonumber \lefteqn{p_{s}^{z}(k|k_{0}; z) = \sum_{t=0}^{\infty}z^{t}p_{s}(k|k_{0}; t)} & \\
\nonumber &= \sum_{m=\left\lceil\frac{\Delta}{2}\right\rceil}^{\infty}{2m\choose\Delta}\left(1-\frac{1}{N}\right)^{2m-\Delta}\frac{1}{N^{\Delta}}\frac{\left(zp\right)^{m}}{\left[1-z\left(1-p\right)\right]^{m+1}}, \\
\label{pz_ER}
\end{align}
from which one can also evaluate $p_{s}^{z}(k|k;z)$ by taking $k_{0}=k$ (or $\Delta=0$). Then, using the relation
\begin{align}
\nonumber \lefteqn{\sum_{m=\left\lceil\frac{\Delta}{2}\right\rceil}^{\infty}{2m\choose\Delta}x^{2m} = \frac{x^{\Delta}}{2}\Big[\left(1-x\right)^{-\Delta-1} +}& \\
\nonumber &+ \left(-1\right)^{\Delta}\left(1+x\right)^{-\Delta-1}\Big]\quad (\Delta\in\mathbb{N},x\in(-1,1)\subset\mathbb{R}), \\
\label{rel1}
\end{align}
which can be seen by combining the expansion of $(1\pm x)^{-\Delta-1}$ for $|x|<1$, it is now possible to obtain the function $f_{s}^{z}(k|k_{0};z)$

\begin{align}
\nonumber f_{s}^{z}(k|k_{0};z) &= \displaystyle\frac{1-\zeta^{2}}{2\left(N-1\right)^{\Delta}}\left[ \frac{\zeta^{\Delta}}{\left(1-\zeta\right)^{\Delta+1}} + \frac{\left(-1\right)^{\Delta}\zeta^{\Delta}}{\left(1+\zeta\right)^{\Delta+1}} \right],\\
\label{fz_ER} 
\end{align}
where
\begin{align}
\zeta := \zeta(z) = \left(1-\frac{1}{N}\right)\sqrt{\frac{zp}{1-z\left(1-p\right)}}.
\label{zeta} 
\end{align}

Expanding \eqref{fz_ER} as in \eqref{exp_fz} is a tedious, but direct procedure. From this operation, the arrival probability can be obtained as being
\begin{align}
\mathcal{A}_{\text{\tiny ER}} = 1 - \frac{1}{2N}\left[1-\frac{\left(-1\right)^{\Delta}}{\left(2N-1\right)^{\Delta}}\right].
\label{A_ER}
\end{align}
Although the dynamics suggests that the vertex $s$ can achieve any larger degree if one waits a sufficiently long time, the probability \eqref{A_ER} is less than one. However, this odd result is a consequence of the growing rule, which allows a vertex to increase its degree by two units by forming a loop. In this case, the targeted degree, $k$, may be surpassed from $k-1$ to $k+1$ without being. For this reason, the arrival probability is not $1$. Nonetheless, if one evaluates
\begin{align}
\nonumber\lefteqn{\sum_{t=0}^{\infty}f_{s}(\text{degree $\geq k$}|k_{0};t) =}& \\
\nonumber &= \displaystyle\sum_{t=0}^{\infty}\Big[f_{s}(k|k_{0};t) + \omega_{\text{\tiny ER}}(k+1|k-1)p_{s}(k-1,t)\Big], \\
\label{A_corrected}
\end{align}
which is a correction to \eqref{A_ER}, the arrival probability is $1$, as expected. Note that the arrival probability \eqref{A_ER} tends to $1$ with the size of the network, which is expected since the loop becomes rare with the number of vertices. One should also note that this result is valid for any positive probability $p$ (the case $p=0$ is trivial), but does not depend explicitly on this parameter. As shown below, this parameter scales the time elapsed until a vertex reaches some degree for the first time, but it does not have any impact on the probability of reaching the pre-established degree (except the trivial case $p=0$, when $\mathcal{A}=0$ for $\Delta>0$).

\begin{figure}
\begin{center}
\includegraphics[width=121.6pt]{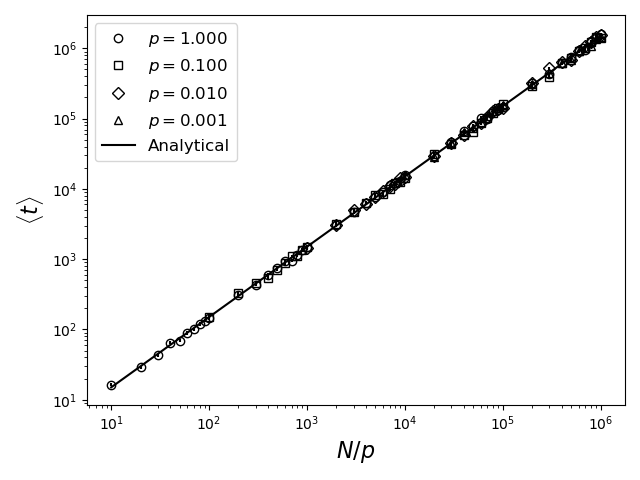}
\includegraphics[width=121.6pt]{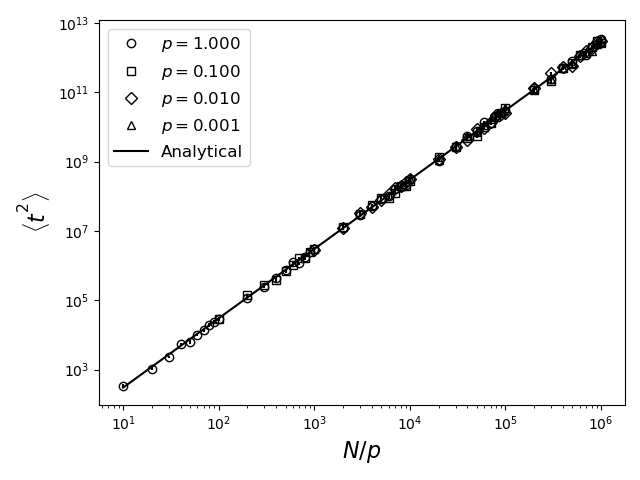}
\caption{\label{fig_er}The mean first (left) and second (right) moments of the first-passage time as a function of the ratio $N/p$ for the dynamical version of the Erd\H os-R\'enyi model with $k_{0}=2$ and $k=5$. The simulations used $100$ samples and compared with the asymptotic results \eqref{<t>_ER} and \eqref{<t2>_ER}; the error bars are smaller than the size of the points.}
\end{center}
\end{figure}

The first and second time moments can also be derived from \eqref{fz_ER}. The leading term of the mean first-passage time is
\begin{align}
\langle t \rangle_{\text{\tiny ER}} \simeq \frac{N\Delta}{2p}
\label{<t>_ER}
\end{align}
for $N\gg 1$, while the second moment is
\begin{align}
\langle t^{2} \rangle_{\text{\tiny ER}} \simeq \left(\frac{N}{2p}\right)^{2}\Delta\left(\Delta+1\right).
\label{<t2>_ER}
\end{align}
The variance can also be determined from \eqref{<t>_ER} and \eqref{<t2>_ER}, and depends quadratically on $N/2p$, but linearly on the difference $\Delta:=k-k_{0}$ as $\sigma^{2}_{\text{\tiny ER}}:=\langle t^{2}\rangle_{\text{\tiny ER}}-\langle t\rangle_{\text{\tiny ER}}^{2}\simeq\left(\frac{N}{2p}\right)^{2}\Delta$. The results \eqref{<t>_ER} and \eqref{<t2>_ER} are supported by numerical simulations, as one can see in figure \eqref{fig_er}.

\subsection{\label{sec:level4B}Time-dependent Watts-Strogatz model}

The analysis of the dynamical version of the Watts-Strogatz model is much more intricate than the previous model. To convey better the ideas, all the technical details are presented in the supplemental material, and we will restrict ourselves to highlighting only the important points in this subsection.

The dynamics of this model was already presented in \eqref{rr_WS}, where the transition rates are given in \eqref{w(k-1)_WS}, \eqref{w(k+1)_WS} and \eqref{w(k)_WS}. Introducing a characteritic function that transforms both the degree and time variables (see \eqref{characteristic_K} and \eqref{characteristic_t}) into new ones, the recurrence relation \eqref{rr_WS} can be converted into the differential equation
\begin{align}
\nonumber \lefteqn{\frac{\partial}{\partial K} p_{s}^{Kz}(K,z) =} & \\
\nonumber &= - \frac{M}{p} \left( \frac{1-z^{-1}}{1-K} + \frac{1-p-z^{-1}}{N+K-1} \right) p_{s}^{Kz}(K,z) - \\
 & - \frac{Mz^{-1}}{p}\left( \frac{1}{1-K} + \frac{1}{N+K-1} \right) p_{s}^{K}(K,t=0).
\label{ode_WS}
\end{align}
Using the normalization condition $p_{s}^{Kz}(K=1,z)=\frac{1}{1-z}$ and assuming $N\gg 1$, the solution of \eqref{ode_WS} can be cast as
\begin{align}
\nonumber p_{s}^{Kz}(K,z) &= \frac{Mz^{-1}}{p}\left(1-K\right)^{-M\alpha}e^{-c\left(1+\alpha\right)\left(1-K\right)} \times \\
  &\times \int_{K}^{1} \text{d}\xi e^{c\left(1+\alpha\right)\left(1-\xi\right)}\left(1-\xi\right)^{M\alpha-1} p_{s}^{z}(\xi,t=0),
\label{sol_ode_WS}
\end{align}
where
\begin{align}
\alpha := \frac{1}{p}\left(z^{-1}-1\right).
\label{alpha}
\end{align}

Then, returning back to the degree variable by inverting the transform \eqref{characteristic_K} leads to
\begin{align}
\nonumber p_{s}^{z}(k|m;z) &= \frac{1}{k!}\frac{\partial^{k}}{\partial K^{k}}\Bigg[ \frac{Mz^{-1}}{p}A_{\alpha}^{-1}(K) \times \\
 &\times \int_{K}^{1}\textup{d}\xi\left(1-\xi\right)^{-1}A_{\alpha}(\xi)\xi^{m} \Bigg]_{K\rightarrow 0},
\label{pkz_WS}
\end{align}
for $m\in\{k,k_{0}\}$ and
\begin{align}
A_{\alpha}(K) := e^{c\left(1+\alpha\right)\left(1-K\right)}\left(1-K\right)^{M\alpha}.
\label{A(K)}
\end{align}

The probability \eqref{pkz_WS} is the key function to compute \eqref{fz_pz}, which can be used to evaluate some quantities of interest through \eqref{exp_fz}. This procedure is not direct as it was in the case of the dynamical Erd\H os-R\'enyi model and the technicalities are exposed in the supplemental material. Here, we will show the results only.

The arrival probability in this model is
\begin{align}
\mathcal{A}_{\text{\tiny WS}} = 1,
\label{A_WS}
\end{align}
as expected. No anomalous behavior as seen in the previous model is present here, where the degree changes by a single unit only.

\begin{figure}
\begin{center}
\includegraphics[width=121.6pt]{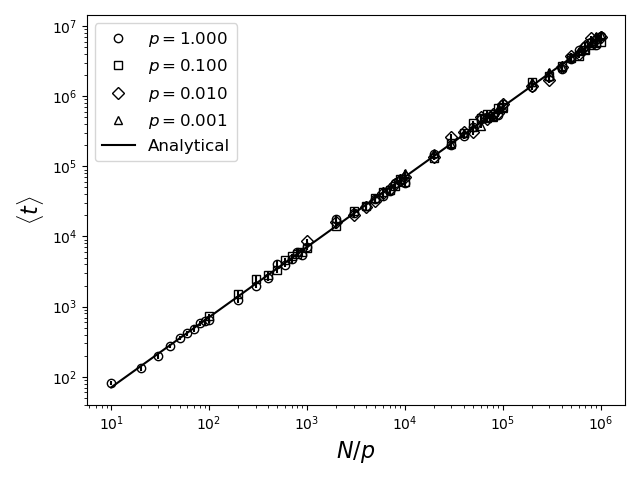}
\includegraphics[width=121.6pt]{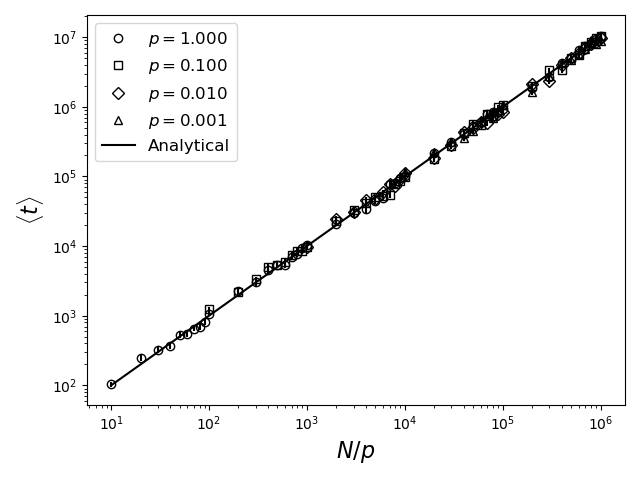}
\caption{\label{fig1}The mean first-passage time as a function of the ratio $N/p$. Left: $k_{0}=2$ and $k=5$; right: $k_{0}=5$ and $k=2$. In both graphs, the mean degree of the network is $c=4$ and the results were obtained from $100$ samples; the error bar is smaller than the size of the points. These simulations were compared with the analytical result \eqref{<t>_WS}.}
\end{center}
\end{figure}

The leading term of the first-passage time is
\begin{align}
\langle t\rangle_{\text{\tiny WS}} \sim \left\{
\begin{array}{ccl}
\displaystyle\frac{N}{p}e^{c}\sum_{n=k_{0}}^{k-1}\frac{\Gamma(n+1,c)}{c^{n}} &,& k>k_{0} \\
 & & \\
\displaystyle\frac{N}{p}e^{c}\sum_{n=k}^{k_{0}-1}\frac{\gamma(n+1,c)}{c^{n}} &,& k<k_{0}
\end{array}
\right.,
\label{<t>_WS}
\end{align}
where $\Gamma(\cdot,\cdot)$ and $\gamma(\cdot,\cdot)$ are, respectively, the upper and lower incomplete Gamma functions. It is worth mentioning that this time is also proportional to $N/p$, as in the dynamical Erd\H os-R\'enyi model. The simulation of this model supports the analytical expression \eqref{<t>_WS}, as shown in figure \ref{fig1}.

On the other hand, the leading contribution to the second moment is given by
\begin{widetext}
\begin{align}
\langle t^{2}\rangle_{\text{\tiny WS}} \sim \left\{
\begin{array}{ccl}
\displaystyle 2\left(\frac{N}{p}\right)^{2}e^{c}\sum_{n=k_{0}}^{k-1}\frac{n!}{c^{n}}\sum_{\ell=0}^{n}\frac{c^{\ell}}{\ell!}\sum_{m=\ell}^{k-1}\frac{\Gamma(m+1,c)}{c^{m}} &,& k>k_{0} \\
 & & \\
\displaystyle 2\left(\frac{N}{p}\right)^{2}e^{c}\sum_{n=k}^{k_{0}-1}\frac{n!}{c^{n}}\sum_{\ell=n+1}^{\infty}\frac{c^{\ell}}{\ell!}\sum_{m=k}^{\ell-1}\frac{\gamma(m+1,c)}{c^{m}} &,& k<k_{0} \\
\end{array}
\right.,
\label{<t2>_WS}
\end{align}
\end{widetext}
and is proportional to $\left(N/p\right)^{2}$. The validity of \eqref{<t2>_WS} was tested by comparing to simulation in figure \ref{fig2}. There is an alternative representation of \eqref{<t2>_WS} in the supplemental material, but the form given here seems to be the most compact one. Naturally, \eqref{<t2>_WS} and \eqref{<t>_WS} can be used to compute the variance, which is also proportional to $\left(N/p\right)^{2}$. Since this expression shows no special aesthetic appeal, it will not be presented here.

\begin{figure}
\begin{center}
\includegraphics[width=121.6pt]{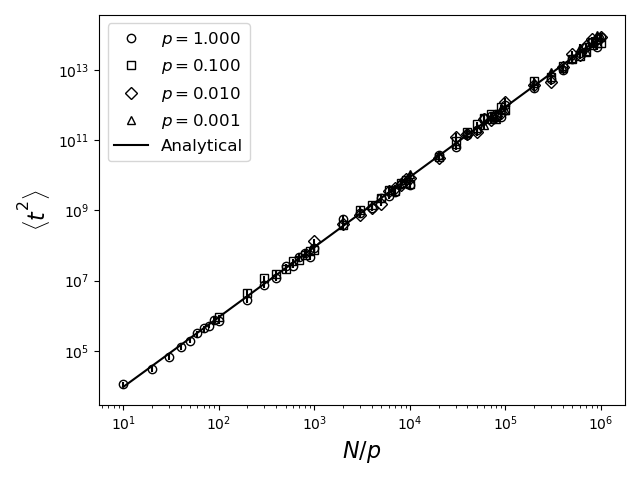}
\includegraphics[width=121.6pt]{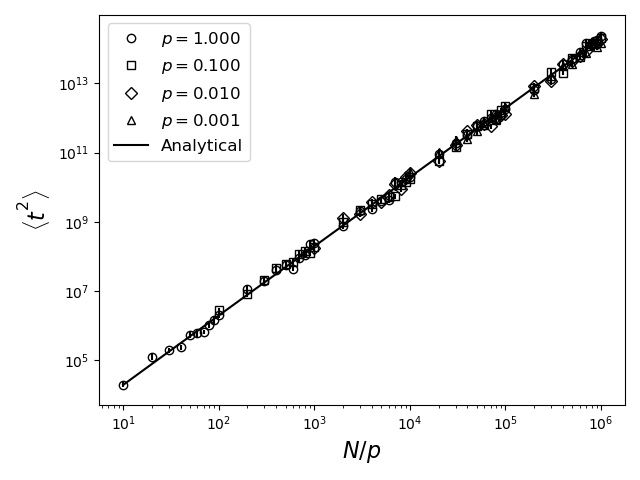}
\caption{\label{fig2}The (mean) second moment of the first-passage time as a function of the ratio $N/p$. Left: $k_{0}=2$ and $k=5$; right: $k_{0}=5$ and $k=2$. In both graphs, the mean degree of the network is $c=4$ and the results were obtained from $100$ samples; the error bar is smaller than the size of the points. These simulations were compared with the analytical result \eqref{<t2>_WS}.}
\end{center}
\end{figure}


\section{\label{conclusion}Conclusion}

In this work, we investigated the time needed for a vertex to achieve a pre-established degree for the first time. The main strategy was mapping the problem into a first-passage problem in degree space. The gain/loss of degrees was illustrated by the time-dependent version of the Erd\H os-R\'enyi and Watts-Strogatz models, which display time-translational symmetry. This property was explored and analytical results concerning the first and second moments of the first-passage time were obtained. In both cases, the arrival probability ensured that the pre-established degree is achieved with probability $1$ (with a careful interpretation in the case of the Erd\H os-R\'enyi dynamics). Furthermore, the mean first-passage time is scaled linearly with the ratio $N/p$ for both models in the asymptotic regime of large networks, while this scale is quadratic for the second moment also in both models. On the other hand, these moments depend on the difference $\Delta$ in the Erd\H os-R\'enyi network only. In the dynamical Watts-Strogatz model, both moments depend on the initial and final degrees independently, as can be seen from \eqref{<t>_WS} and \eqref{<t2>_WS}, although the structure of both $\langle t\rangle_{\text{\tiny WS}}$ and $\langle t^{2}\rangle_{\text{\tiny WS}}$ can be compactly represented as a sum of terms involving upper (lower) incomplete Gamma functions when the final degree is larger (smaller) than the initial one.


\section*{\label{acknowledgements}Acknowledgments}

F.A. was financed in part by the Coordena\c{c}\~ao de Aperfei\c{c}oamento de Pessoal de N\'ivel Superior - Brasil (CAPES) - Finance Code 001.

\pagebreak
\onecolumngrid

\setcounter{equation}{0}
\setcounter{figure}{0}
\setcounter{table}{0}
\setcounter{page}{1}
\renewcommand{\theequation}{S\arabic{equation}}
\renewcommand{\thefigure}{S\arabic{figure}}
\renewcommand{\bibnumfmt}[1]{[S#1]}
\renewcommand{\citenumfont}[1]{S#1}

\section{Supplemental material}

The next sections provide the details to obtain the first and second moments of the first-passage distribution of the dynamical Watts-Strogatz model. For convenience, some definitions and results from the main text are repeated here.


\section{First-passage function}

The starting point is the recurrence relation
\begin{align}
p_{s}(k,t+1) = \omega_{\text{\tiny WS}}(k|k-1)p_{s}(k-1,t) + \omega_{\text{\tiny WS}}(k|k+1)p_{s}(k+1,t) + \omega_{\text{\tiny WS}}(k|k)p_{s}(k,t),
\label{rec_ws}
\end{align}
where $p_{s}(k,t)$ is the probability of the vertex $s$ having degree $k$ at time $t$ and $\omega_{\text{\tiny WS}}(k|m)$ is the conditional probability of a vertex of degree $m$ change to a state of degree $k$. For the dynamical version of the Watts-Strogatz model, these probabilities are
\begin{align}
\omega_{\text{\tiny WS}}(k|k-1) = \left(1-\frac{k-1}{M}\right)\frac{p}{N},\quad \omega_{\text{\tiny WS}}(k|k+1) = \frac{k+1}{M}\left(1-\frac{1}{N}\right)p,\quad\textnormal{ and }\quad\omega_{\text{\tiny WS}}(k|k) = 1 + p\left(\frac{2k}{MN} - \frac{1}{N} - \frac{k}{M}\right),
\label{w_ws}
\end{align}
as shown in the main text. Here, $p$ is the linking probability, and $N$ and $M$ are the total number of vertices and degrees, respectively, related by
\begin{align}
M = cN,
\label{M=k0N}
\end{align}
where $c$ is the mean degree of the Watts-Strogatz network. From the discrete Laplace transforms
\begin{align}
p_{s}^{K}(K,t) = \sum_{k=0}^{\infty}K^{k}p_{s}(k,t) \quad\textnormal{ and }\quad p_{s}^{Kz}(K,z) = \sum_{t=0}^{\infty}z^{t}p_{s}^{K}(K,t),
\label{laplacetransform}
\end{align}
the recurrence relation \eqref{rec_ws} can be cast as
\begin{align}
\partial_{K}p_{s}^{Kz}(K,t) = -\frac{M}{p}\left(\frac{1-z^{-1}}{1-K} + \frac{1-p-z^{-1}}{N+K-1}\right)p_{s}^{Kz}(K,z) - \frac{Mz^{-1}}{p}\left(\frac{1}{1-K} + \frac{1}{N+K-1}\right)p_{s}^{K}(K,t=0).
\label{pde_PKz}
\end{align}
Using the boundary condition $p_{s}^{Kz}(1,z)=(1-z)^{-1}$, which is the normalization condition, the solution of the differential equation \eqref{pde_PKz} is
\begin{align}
p_{s}^{Kz}(K,t) = \frac{MNz^{-1}}{p}\frac{\left(N+K-1\right)^{M\left(\alpha+1\right)}}{\left(1-K\right)^{M\alpha}}\int_{K}^{1}\textup{d}\xi\,\frac{\left(1-\xi\right)^{M\alpha-1}}{\left(N+\xi-1\right)^{M\alpha+M+1}}p_{s}^{K}(\xi,t=0),
\label{solution_pde_PKz}
\end{align}
where
\begin{align}
\alpha := \frac{1}{p}\left(z^{-1}-1\right).
\label{alpha}
\end{align}
Since
\begin{align}
\left(N+K-1\right)^{M\alpha+M} = N^{M\left(\alpha+1\right)}e^{-c\left(\alpha+1\right)\left(1-K\right)} \left[1 + \mathcal{O}(N^{-1})\right],
\label{exponentialapproximation}
\end{align}
the Laplace-transformed degree distribution \eqref{solution_pde_PKz} equals
\begin{align}
p_{s}^{Kz}(K,t) = \frac{Mz^{-1}}{p}\left(1-K\right)^{-M\alpha}e^{-c\left(\alpha+1\right)\left(1-K\right)}\int_{K}^{1}\textup{d}\xi\,\left(1-\xi\right)^{M\alpha-1}e^{c\left(\alpha+1\right)\left(1-\xi\right)}p_{s}^{K}(\xi,t=0),
\label{solution_pde}
\end{align}
except for terms which are $\mathcal{O}(N^{-1})$ times smaller.

Defining
\begin{align}
A_{\alpha}(K) := e^{c\left(1+\alpha\right)\left(1-K\right)}\left(1-K\right)^{M\alpha},
\label{A}
\end{align}
and taking the inverse Laplace transformation in the first argument of $P^{Kz}(K,z)$ through
\begin{align}
p_{s}^{z}(k,z) = \frac{1}{k!}\lim_{K\rightarrow 0}\frac{\textup{d}^{k}}{\textup{d}K^{k}}p_{s}^{Kz}(K,z)
\label{invLaplaceK}
\end{align}
leads to
\begin{align}
p_{s}^{z}(k,z) = \frac{1}{k!}\lim_{K\rightarrow 0}\frac{\textup{d}^{k}}{\textup{d}K^{k}}\left[\frac{Mz^{-1}}{p}A_{\alpha}^{-1}(K)\int_{K}^{1}\textup{d}\xi\,\left(1-\xi\right)^{-1}A_{\alpha}(\xi)P^{K}(\xi,t=0)\right].
\label{prePz}
\end{align}
One should keep in mind that the expression above for $p_{s}^{z}$ denpends on the initial condition of $p_{s}^{K}$ at time $t=0$. If the network starts with degree $m$, one has $p_{s}^{K}(\xi,t=0)=\xi^{m}$, which implies
\begin{align}
p_{s}^{z}(k|m;z) = \frac{1}{k!}\lim_{K\rightarrow 0}\frac{\textup{d}^{k}}{\textup{d}K^{k}}\left[\frac{Mz^{-1}}{p}A_{\alpha}^{-1}(K)\int_{K}^{1}\textup{d}\xi\,\left(1-\xi\right)^{-1}A_{\alpha}(\xi)\xi^{m}\right],
\label{Pz}
\end{align}
and a new notation, $p_{s}^{z}(k|m;z)$, was introduced to make clear the initial degree. Then, the key function $f_{s}^{z}$ defined in the main text can be cast as
\begin{align}
f_{s}^{z}(k|k_{0};z) = \frac{p_{s}^{z}(k|k_{0};z)}{p_{s}^{z}(k|k;z)} = \frac{\frac{\textup{d}^{k}}{\textup{d}K^{k}}\left[A_{\alpha}^{-1}(K)\int_{K}^{1}\textup{d}\xi\,\left(1-\xi\right)^{-1}A_{\alpha}(\xi)\xi^{k_{0}}\right]_{K\rightarrow 0}}{\frac{\textup{d}^{k}}{\textup{d}K^{k}}\left[A_{\alpha}^{-1}(K)\int_{K}^{1}\textup{d}\xi\,\left(1-\xi\right)^{-1}A_{\alpha}(\xi)\xi^{k}\right]_{K\rightarrow 0}}.
\label{Fz}
\end{align}
The main point investigated is the following: given a vertex of the network (with $N$ vertices and total degree $M=cN$, where $c$ is the mean degree) with degree $k_{0}$, one is interested in the time needed to it reach degree $k$ for the first time.

For $\lambda\in\{k_{0},k\}$, and expanding the multiple derivatives in \eqref{Fz} leads to
\begin{align}
\nonumber\frac{\textup{d}^{k}}{\textup{d}K^{k}}\left[A_{\alpha}^{-1}(K)\int_{K}^{1}\textup{d}\xi\,\left(1-\xi\right)^{-1}A_{\alpha}(\xi)\xi^{\lambda}\right]_{K\rightarrow 0} &= \left[A_{\alpha}^{-1}(K)\right]_{K\rightarrow 0}^{(k)}\int_{0}^{1}\textup{d}\xi\,\left(1-\xi\right)^{-1}A_{\alpha}(\xi)\xi^{\lambda} + \\
  &+\theta(k-1)\sum_{\ell=1}^{k}{k\choose\ell}\left[A_{\alpha}^{-1}(K)\right]_{K\rightarrow 0}^{(k-\ell)}\left[\int_{K}^{1}\textup{d}\xi\,\left(1-\xi\right)^{-1}A_{\alpha}(\xi)\xi^{\lambda}\right]_{K\rightarrow 0}^{(\ell)},
\label{expansiondK}
\end{align}
where $f^{(n)}$ stands for the $n$th derivative of $f$ and
\begin{align}
\theta(x)=\left\{
\begin{array}{ccl}
1 &,& x\geq 0 \\
0 &,& x<0
\end{array}
\right.
\label{unitstep}
\end{align}  
is the unit step funtion. Although the introduction of this function is redundant (for instance, one might define $\sum_{\ell=1}^{0}(\cdots)$ to be zero in \eqref{expansiondK}), it helps to keep in mind the range of the variables that produces nonzero contributions. This is particularly useful in later calculations where many of such constraints appear.

One should note that
\begin{align}
\nonumber\left[\int_{K}^{1}\textup{d}\xi\,\left(1-\xi\right)^{-1}A_{\alpha}(\xi)\xi^{\lambda}\right]_{K\rightarrow 0}^{(\ell)} &= \left[-\left(1-K\right)^{-1}A_{\alpha}(K)K^{\lambda}\right]_{K\rightarrow 0}^{(\ell-1)} \\
 &=\left\{
\begin{array}{ccl}
0 &,& \lambda > \ell-1 \\
 & & \\
-\lambda!\theta(\ell-1-\lambda){\ell-1\choose\lambda}\left[\left(1-K\right)^{-1}A_{\alpha}(K)\right]_{K\rightarrow 0}^{(\ell-1-\lambda)} &,& \lambda\leq\ell-1
\end{array}
\right.,
\label{k=0k0neq0}
\end{align}
where the factor $\theta(\ell-1-\lambda)$ is a reminder that the condition $\lambda\leq\ell-1$ should be satisfied to generate a nonzero result.

From \eqref{k=0k0neq0}, the expression \eqref{Fz} can be written as
\begin{align}
\nonumber f_{s}^{z}(k|k_{0}; z) &= \frac{\int_{0}^{1}\textup{d}\xi\,\left(1-\xi\right)^{-1}A_{\alpha}(\xi)\xi^{k_{0}}}{\int_{0}^{1}\textup{d}\xi\,\left(1-\xi\right)^{-1}A_{\alpha}(\xi)\xi^{k}} - \\
\nonumber  &- \theta(k-1)\frac{\sum_{\ell=1}^{k}{k\choose\ell}\left[A_{\alpha}^{-1}(K)\right]_{K\rightarrow 0}^{(k-\ell)}k_{0}!\theta(\ell-1-k_{0}){\ell-1\choose k_{0}}\left[\left(1-K\right)^{-1}A_{\alpha}(K)\right]_{K\rightarrow 0}^{(\ell-1-k_{0})} }{\left[A_{\alpha}^{-1}(K)\right]_{K\rightarrow 0}^{(k)}\int_{0}^{1}\textup{d}\xi\,\left(1-\xi\right)^{-1}A_{\alpha}(\xi)\xi^{k}} \\
\nonumber &= \frac{\int_{0}^{1}\textup{d}\xi\,\left(1-\xi\right)^{-1}A_{\alpha}(\xi)\xi^{k_{0}}}{\int_{0}^{1}\textup{d}\xi\,\left(1-\xi\right)^{-1}A_{\alpha}(\xi)\xi^{k}} - \\
 &- \theta(k-1)\theta(k-k_{0}-1)\frac{\sum_{\ell=k_{0}+1}^{k}{k\choose\ell}\left[A_{\alpha}^{-1}(K)\right]_{K\rightarrow 0}^{(k-\ell)}k_{0}!{\ell-1\choose k_{0}}\left[\left(1-K\right)^{-1}A_{\alpha}(K)\right]_{K\rightarrow 0}^{(\ell-1-k_{0})} }{\left[A_{\alpha}^{-1}(K)\right]_{K\rightarrow 0}^{(k)}\int_{0}^{1}\textup{d}\xi\,\left(1-\xi\right)^{-1}A_{\alpha}(\xi)\xi^{k}}.
\label{Fz1}
\end{align}
From now on, the notation
\begin{align}
\Delta := k - k_{0}
\label{Delta}
\end{align}
will be adopted. Therefore, from $\theta(k-1)\theta(k-k_{0}-1) = \theta(k-k_{0}-1) = \theta(\Delta-1)$, one has
\begin{align}
\nonumber f_{s}^{z}(k|k_{0}; z) &= \frac{\int_{0}^{1}\textup{d}\xi\,\left(1-\xi\right)^{-1}A_{\alpha}(\xi)\xi^{k_{0}}}{\int_{0}^{1}\textup{d}\xi\,\left(1-\xi\right)^{-1}A_{\alpha}(\xi)\xi^{k}} - \\
 &- \theta(\Delta-1)\frac{\sum_{\ell=k_{0}+1}^{k}{k\choose\ell}{\ell-1\choose k_{0}}k_{0}!\left[A_{\alpha}^{-1}(K)\right]_{K\rightarrow 0}^{(k-\ell)}\left[\left(1-K\right)^{-1}A_{\alpha}(K)\right]_{K\rightarrow 0}^{(\ell-1-k_{0})} }{\left[A_{\alpha}^{-1}(K)\right]_{K\rightarrow 0}^{(k)}\int_{0}^{1}\textup{d}\xi\,\left(1-\xi\right)^{-1}A_{\alpha}(\xi)\xi^{k}}.
\label{Fz2}
\end{align}


\section{Some key functions}

In this section, the behavior of the following functions will be examined.
\begin{enumerate}

\item $\int_{0}^{1}\textup{d}\xi\,\left(1-\xi\right)^{-1}A_{\alpha}(\xi)\xi^{\lambda}$, $\lambda\in\{k_{0},k\}$
  
\item $\left[A_{\alpha}^{-1}(K)\right]_{K\rightarrow 0}^{(n)}$

\item $\left[\left(1-K\right)^{-1}A_{\alpha}(K)\right]_{K\rightarrow 0}^{(n)}$
  
\end{enumerate}
The asymptotic behavior of these three functions for small $\alpha$ will be considered. The first and second-order terms in $\alpha$ are directly related to the first and second first-passage time moments.


\subsection{The function $\int_{0}^{1}\textup{d}\xi\,\left(1-\xi\right)^{-1}A_{\alpha}(\xi)\xi^{\lambda}$, $\lambda\in\{k_{0},k\}$}

Integrating $\int_{0}^{1}\textup{d}\xi\,\left(1-\xi\right)^{-1}A_{\alpha}(\xi)\xi^{\lambda}$ ($\lambda\in\{k_{0},k\}$ and $A_{\alpha}$ given by \eqref{A}) by parts yields
\begin{align}
\int_{0}^{1}\textup{d}\xi\,\left(1-\xi\right)^{-1}A_{\alpha}(\xi)\xi^{\lambda} = \frac{1}{M\alpha}\left\{ \delta_{\lambda,0}e^{c\left(1+\alpha\right)} + \int_{0}^{1}\textup{d}\xi\,\left(1-\xi\right)^{M\alpha}e^{c\left(1+\alpha\right)\left(1-\xi\right)}\xi^{\lambda-1}\left[\lambda-c\xi\left(1+\alpha\right)\right] \right\},
\label{f1_1}
\end{align}
where $\delta_{m,n}$ stands for the Kronecker delta, defined by
\begin{align}
\delta_{m,n} = \left\{
\begin{array}{ccl}
1 &,& m=n \\
0 &,& m\neq n
\end{array}
\right..
\label{kronoecker}
\end{align}
By expanding the expression inside the brace in \eqref{f1_1} for $\alpha\sim 0$ leads to
\begin{align}
\nonumber\lefteqn{\int_{0}^{1}\textup{d}\xi\,\left(1-\xi\right)^{-1}A_{\alpha}(\xi)\xi^{\lambda} =}&\\
\nonumber  &= \frac{1}{M\alpha}\Bigg\{ e^{c}\delta_{\lambda,0} + \int_{0}^{1}\textup{d}\xi\,e^{c\left(1-\xi\right)}\xi^{\lambda-1}\left(\lambda-c\xi\right) + \\
\nonumber &+ \alpha ce^{c}\delta_{\lambda,0} + \alpha\int_{0}^{1}\textup{d}\xi\,e^{c\left(1-\xi\right)}\xi^{\lambda-1}\left[M\left(\lambda-c\xi\right)\ln\left(1-\xi\right)+c\left(\lambda-c\xi\right)\left(1-\xi\right) - c\xi \right] + \\
\nonumber &+ \alpha^{2}\frac{c^{2}e^{c}}{2}\delta_{\lambda,0} + \alpha^{2}\int_{0}^{1}\textup{d}\xi\,e^{c\left(1-\xi\right)}\xi^{\lambda-1}\Bigg[ \frac{M^{2}}{2}\left(\lambda-c\xi\right)\ln^{2}\left(1-\xi\right) + Mc\left(\lambda-c\xi\right)\left(1-\xi\right)\ln\left(1-\xi\right) - \\
  &- Mc\xi\ln\left(1-\xi\right) + \frac{c^{2}}{2}\left(\lambda-c\xi\right)\left(1-\xi\right)^{2} - c^{2}\xi\left(1-\xi\right) \Bigg] + \mathcal{O}(\alpha^{3})\Bigg\}.
\label{f1_2}
\end{align}

Defining
\begin{align}
\left\{
\begin{array}{ccl}
\Omega_{1}(\lambda) &:=& \displaystyle\int_{0}^{1}\textup{d}\xi\,e^{c\left(1-\xi\right)}\xi^{\lambda-1}\left(\lambda-c\xi\right)\ln\left(1-\xi\right) \\
 & & \\
\Omega_{2}(\lambda) &:=& \displaystyle\int_{0}^{1}\textup{d}\xi\,e^{c\left(1-\xi\right)}\xi^{\lambda-1}\left(\lambda-c\xi\right)\ln^{2}\left(1-\xi\right)
\end{array}
\right.
\quad (n\in\{1,2\})
\label{Omega12}
\end{align}
and from the relation
\begin{align}
\int_{0}^{1}\textup{d}\xi\,e^{c\left(1-\xi\right)}\xi^{m} = m!c^{-m-1}\left(e^{c}-\sum_{\ell=0}^{m}\frac{c^{\ell}}{\ell !}\right) = m!c^{-m-1}\sum_{\ell=m+1}^{\infty}\frac{c^{\ell}}{\ell !}, \quad (m\in\{0,1,2,\ldots\}),
\label{expmon}
\end{align}
one can cast \eqref{f1_2} as
\begin{align}
\nonumber\int_{0}^{1}\textup{d}\xi\,\left(1-\xi\right)^{-1}A_{\alpha}(\xi)\xi^{\lambda} &= \frac{1}{M\alpha}\Bigg[ 1 + \alpha M\Omega_{1}(\lambda) + \frac{\alpha^{2}M^{2}}{2}\Omega_{2}(\lambda) + \alpha^{2}Mc\int_{0}^{1}\textup{d}\xi\,e^{c\left(1-\xi\right)}\xi^{\lambda-1}\left(\lambda-c\xi\right)\left(1-\xi\right)\ln\left(1-\xi\right) - \\
 &- \alpha^{2}Mc\int_{0}^{1}\textup{d}\xi\,e^{c\left(1-\xi\right)}\xi^{\lambda}\ln\left(1-\xi\right) + \mathcal{O}(\alpha^{3})\Bigg].
\label{f1_3}
\end{align}
after some lengthy (but direct) calculations. Integrating the last integral in \eqref{f1_3}, $\int_{0}^{1}\textup{d}\xi\,e^{c\left(1-\xi\right)}\xi^{\lambda}\ln\left(1-\xi\right)$, by parts (choosing the resulting integral to be composed by the antiderivative of $\ln\left(1-\xi\right)$ and derivative of $e^{c\left(1-\xi\right)}\xi^{\lambda}$), the term associated to $\alpha^{2}$ is simplified. Using \eqref{expmon} again, one can finally state that
\begin{align}
\int_{0}^{1}\textup{d}\xi\,\left(1-\xi\right)^{-1}A_{\alpha}(\xi)\xi^{\lambda} = \frac{1}{M\alpha}\Bigg\{ 1 + \alpha M\Omega_{1}(\lambda) + \frac{\alpha^{2}M^{2}}{2}\Omega_{2}(\lambda) + \alpha^{2}Mc\int_{0}^{1}\textup{d}\xi\,e^{c\left(1-\xi\right)}\xi^{\lambda-1} + \mathcal{O}(\alpha^{3})\Bigg\}.
\label{(i)}
\end{align}


\subsection{The function $\left[A_{\alpha}^{-1}(K)\right]_{K\rightarrow 0}^{(n)}$}

From the definition \eqref{A} of $A_{\alpha}$,
\begin{align}
\nonumber\left[A_{\alpha}^{-1}(K)\right]_{K\rightarrow 0}^{(n)} &= \sum_{\ell=0}^{n}{n\choose\ell}\Big[e^{-c\left(1+\alpha\right)\left(1-K\right)}\Big]_{K\rightarrow 0}^{(n-\ell)}\Big[\left(1-K\right)^{-M\alpha}\Big]_{K\rightarrow 0}^{(\ell)} \\
\nonumber &= \sum_{\ell=0}^{n}{n\choose\ell}e^{-c\left(1+\alpha\right)}\left[c\left(1+\alpha\right)\right]^{n-\ell}\frac{\Gamma(M\alpha+\ell)}{\Gamma(M\alpha)} \\
 &= \left[c\left(1+\alpha\right)\right]^{n}e^{-c\left(1+\alpha\right)} + \theta(n-1)\sum_{\ell=1}^{n}{n\choose\ell}e^{-c\left(1+\alpha\right)}\left[c\left(1+\alpha\right)\right]^{n-\ell}\left(M\alpha\right)\cdots\left(M\alpha+\ell-1\right),
\label{A-1derivative0}
\end{align}
where the sum was splitted in the $\ell=0$ term and the $\ell>0$ ones. Again, the factor $\theta(n-1)$ was introduced as a reminder that the last sum is nonzero for $n\geq 1$ only. Expanding \eqref{A-1derivative0} for small $\alpha$ leads to
\begin{align}
\left[A_{\alpha}^{-1}(K)\right]_{K\rightarrow 0}^{(n)} = c^{n}e^{-c} + \alpha\left(n-c\right)c^{n}e^{-c} + \alpha Me^{-c}\theta(n-1)n!\sum_{\ell=0}^{n-1}\frac{c^{\ell}}{\ell!}\frac{1}{n-\ell} + \mathcal{O}(\alpha^{2}).
\label{(ii)}
\end{align}
There is no need to determine exactly the second-order term in this case, as it will be seen later.


\subsection{The function $\left[\left(1-K\right)^{-1}A_{\alpha}(K)\right]_{K\rightarrow 0}^{(n)}$}

One may first expand $\left(1-K\right)^{-1}A_{\alpha}(K)$ in power of $\alpha$ first, which leads to
\begin{align}
\left[\left(1-K\right)^{-1}A_{\alpha}(K)\right]_{K\rightarrow 0}^{(n)} = \left[ e^{c\left(1-K\right)}\left(1-K\right)^{-1} + \alpha ce^{c\left(1-K\right)} + \alpha Me^{c\left(1-K\right)}\left(1-K\right)^{-1}\ln\left(1-K\right) \right]_{K\rightarrow 0}^{(n)} + \mathcal{O}(\alpha^{2}).
\label{(iii)-0}
\end{align}
Then, from
\begin{align}
\left[e^{c\left(1-K\right)}\left(1-K\right)^{-1}\right]_{K\rightarrow 0}^{(n)} = \sum_{\ell=0}^{n}{n\choose\ell}\left[e^{c\left(1-K\right)}\right]_{K\rightarrow 0}^{(n-\ell)}\left[\left(1-K\right)^{-1}\right]_{K\rightarrow 0}^{(\ell)} = n!e^{c}\sum_{u=0}^{n}\frac{\left(-c\right)^{u}}{u!}
\label{(iii)-1}
\end{align}
and $\left[\ln\left(1-K\right)\right]_{K\rightarrow 0}^{(r)}=-\theta(r-1)(r-1)!$, one has
\begin{align}
\left[\left(1-K\right)^{-1}A_{\alpha}(K)\right]_{K\rightarrow 0}^{(n)} = n!e^{c}\sum_{\ell=0}^{n}\frac{\left(-c\right)^{\ell}}{\ell!} + \alpha\left[ ce^{c}\left(-c\right)^{n} - Mn!e^{c}\theta(n-1)\sum_{\ell=0}^{n-1}\frac{1}{n-\ell}\sum_{m=0}^{\ell}\frac{\left(-c\right)^{m}}{m!} \right] + \mathcal{O}(\alpha^{2}).
\label{(iii)}
\end{align}


\section{Expansion of the first-passage function}

One should now insert \eqref{(i)}, \eqref{(ii)} and \eqref{(iii)} into \eqref{Fz2}, and expanding the resulting expression in $\alpha$. After a lengthy (but direct) calculation, it leads to
\begin{align}
f_{s}^{z}(k|k_{0};z) = 1 + \alpha L(k,k_{0}) + \alpha^{2}Q(k,k_{0}) + \mathcal{O}(\alpha^{3}),
\end{align}
where the coeficients of the linear and quadratic terms are given, respectively, by
\begin{align}
\nonumber L(k,k_{0}) &:= M\left[\Omega_{1}(k_{0})-\Omega_{1}(k)\right] - M\theta(\Delta-1)e^{c}\sum_{\ell=k_{0}+1}^{k}\frac{k!}{\left(k-\ell\right)!}\frac{c^{-\ell}}{\ell}\sum_{m=0}^{\ell-1-k_{0}}\frac{\left(-c\right)^{m}}{m!} \\
 &= M\left[\Omega_{1}(k_{0})-\Omega_{1}(k)\right] - M\theta(\Delta-1)k!e^{c}\sum_{\ell=0}^{\Delta-1}\frac{c^{\ell-k}}{\ell!}\frac{1}{k-\ell}\sum_{m=0}^{\Delta-1-\ell}\frac{\left(-c\right)^{m}}{m!},
\label{L}
\end{align}
where the change of variable $\ell\rightarrow k-\ell$ was performed in the last passage, and
\begin{align}
\nonumber Q(k,k_{0}) &:= M^{2}\left[\frac{1}{2}\Omega_{2}(k_{0})-\frac{1}{2}\Omega_{2}(k)+\Omega_{1}^{2}(k)-\Omega_{1}(k_{0})\Omega_{1}(k)\right] + Mc\int_{0}^{1}\textup{d}\xi\,e^{c\left(1-\xi\right)}\left(\xi^{k_{0}}-\xi^{k}\right) + \\
\nonumber &+ Me^{c}c^{-k}\theta(\Delta-1)\sum_{\ell=k_{0}+1}^{k} \frac{k!}{\left(k-\ell\right)!\ell} \Bigg\{ \frac{c^{k-\ell}\left(-c\right)^{\ell-k_{0}}}{\left(\ell-1-k_{0}\right)!} + \ell c^{k-\ell}\sum_{m=0}^{\ell-1-k_{0}}\frac{\left(-c\right)^{m}}{m!} + \\
\nonumber & + M\Omega_{1}(k)c^{k-\ell}\sum_{m=0}^{\ell-1-k_{0}}\frac{\left(-c\right)^{m}}{m!} - M\left(k-\ell\right)!\theta(k-\ell-1)\sum_{m=0}^{\ell-1-k_{0}}\frac{\left(-c\right)^{m}}{m!}\sum_{q=0}^{k-\ell-1}\frac{c^{q}}{q!}\frac{1}{k-\ell-q} + \\
 &+ Mc^{k-\ell}\theta(\ell-2-k_{0})\sum_{m=0}^{\ell-2-k_{0}}\frac{1}{\ell-1-k_{0}-m}\sum_{q=0}^{m}\frac{\left(-c\right)^{q}}{q!} + Mc^{-\ell}k!\theta(k-1)\sum_{m=0}^{\ell-1-k_{0}}\frac{\left(-c\right)^{m}}{m!}\sum_{q=0}^{k-1}\frac{c^{q}}{q!}\frac{1}{k-q} \Bigg\}.
\label{Q}
\end{align}
 
From now on, the behavior of the coefficients $L(k,k_{0})$ and $Q(k,k_{0})$ is analyzed. In the next section, some useful relations are presented to deal with this task.


\section{Some useful relations}

For any positive $u>0$, the relations
\begin{align}
\frac{1}{u} = \int_{1}^{\infty}\textup{d}x\,x^{-u-1}
\label{1/u}
\end{align}
and
\begin{align}
\frac{1}{u^{\alpha}} = \frac{1}{\Gamma(\alpha)}\int_{0}^{\infty}\textup{d}x\,e^{-ux}x^{\alpha-1} \qquad (\alpha>0)
\label{1/ualpha}
\end{align}
will be used extensively, as well as the relations
\begin{align}
\sum_{n=0}^{a}\frac{x^{n}}{n!} = \frac{e^{x}\Gamma(a+1,x)}{a!} \quad\textnormal{ and }\quad \sum_{n=a+1}^{\infty}\frac{x^{n}}{n!} = \frac{e^{x}\gamma(a+1,x)}{a!},
\label{sumgamma}
\end{align}
where $a$ is a non-negative integer number and $x\in\mathbb{R}$. Furthermore, $\Gamma$ and $\gamma$ are the (upper and lower, respectively) incomplete Gamma functions, defined as
\begin{align}
\Gamma(n,x) := \int_{x}^{\infty}\textup{d}t\,e^{-t}t^{n-1} \quad\textnormal{ and }\quad\gamma(n,x) := \int_{0}^{x}\textup{d}t\,e^{-t}t^{n-1},
\label{incgamma}
\end{align}
and satisfy
\begin{align}
\Gamma(n,x) + \gamma(n,x) = \Gamma(n).
\label{Gg}
\end{align}
Both expressions in \eqref{sumgamma} can be obtained by integrating the incomplete Gamma functions by part repeatedly. The incomplete Gamma functions also satisfy
\begin{align}
\Gamma(n+1,x) = n\Gamma(n,x) + x^{n}e^{-x} \quad\textnormal{ and }\quad \gamma(n+1,x) = n\gamma(n,x) - x^{n}e^{-x}.
\label{recurrencegamma}
\end{align}

Defining
\begin{align}
\psi(\lambda) := \int_{0}^{1}\textup{d}\xi\,e^{c\left(1-\xi\right)}\xi^{\lambda}\ln\left(1-\xi\right)
\label{psi}
\end{align}
for non-negative integer $\lambda$, one can see, from the definition \eqref{Omega12}, that
\begin{align}
\Omega_{1}(\lambda) = \lambda\psi(\lambda-1) - c\psi(\lambda),
\label{Omega1psi}
\end{align}
where the arguments are taken to be non-negative. Inverting this recurrence relation yields
\begin{align}
-\frac{c^{\lambda+1}}{\lambda!}\psi(\lambda) = \sum_{n=0}^{\lambda}\frac{c^{n}}{n!}\Omega_{1}(n).
\label{psiOmega1}
\end{align}
By letting $\lambda\rightarrow\infty$, \eqref{psiOmega1} implies
\begin{align}
\sum_{n=0}^{\infty}\frac{c^{n}}{n!}\Omega_{1}(n) = 0.
\label{sumOmega1=0}
\end{align}
Another important relation is obtained by Taylor-expanding the logarithmic term in \eqref{Omega12} and invoking \eqref{expmon}, which allows one to derive
\begin{align}
\Omega_{1}(a)-\Omega_{1}(b) = e^{c}\sum_{n=a}^{b-1}\frac{\gamma(n+1,c)}{c^{n+1}} \qquad (\textnormal{$a,b\in\{0,1,2,\ldots\}\subset\mathbb{Z}$ and $a<b$}).
\label{Omega1-Omega1}
\end{align}
There is also a relation involving differences of $\Omega_{2}$ functions. Firstly, expanding both logarithms in $\Omega_{2}$ defined in \eqref{Omega12} and using \eqref{expmon}, \eqref{sumgamma} and \eqref{incgamma} yields
\begin{align}
\Omega_{2}(\lambda) = \sum_{n,m=1}^{\infty}\left[\frac{1}{nm} - \left(\frac{n+m}{nm}\right)\int_{0}^{1}\textup{d}u\,e^{c\left(1-u\right)}u^{\lambda+n+m-1}\right].
\label{Omega2_lambda}
\end{align}
Note that the expression inside the square brackets is symmetric with respect to $n$ and $m$, but it is not possible to split into two terms because the sum over each term diverges. Let $a,b\in\{0,1,2,\ldots\}\subset\mathbb{Z}$ with $a<b$. Taking into account the symmetry between $n$ and $m$ in the sum \eqref{Omega2_lambda}, one has
\begin{align}
\frac{1}{2}\left[\Omega_{2}(a) - \Omega_{2}(b)\right] = \sum_{n,m=1}^{\infty}\frac{1}{n}\int_{0}^{1}\textup{d}u\,e^{c\left(1-u\right)}u^{n+m-1}\left(u^{b}-u^{a}\right) = \int_{0}^{1}\textup{d}u\,e^{c\left(1-u\right)}\left(\frac{u^{b}-u^{a}}{1-u}\right)\ln\left(1-u\right) \qquad(a<b).
\label{preOmega2-Omega2}
\end{align}
This formula can also be cast as
\begin{align}
\frac{1}{2}\left[\Omega_{2}(a) - \Omega_{2}(b)\right] = \sum_{n=a}^{b-1}\int_{0}^{1}\textup{d}u\,e^{c\left(1-u\right)}u^{n}\ln\left(1-u\right) = \sum_{n=a}^{b-1}\psi(n),
\label{Omega2-Omega2}
\end{align}
where $a<b$ and $\psi$ is defined in \eqref{psi}.

In the next sections, the following integral is recurrent:
\begin{align}
\int_{0}^{\infty}\textup{d}x\,e^{-\alpha x}x^{\beta}\int_{0}^{\infty}\textup{d}y\,e^{y\left(1-x\right)}\int_{y}^{\infty}\textup{d}z\,e^{-z}z^{\gamma} \qquad (\alpha,\beta,\gamma>0\textnormal{ and }\beta>\gamma).
\label{triple}
\end{align}
This integral shows an \textquotedblleft apparent singularity\textquotedblright, which can be circumvented by a suitable procedure. Splitting the integral into two parts,
\begin{align}
\nonumber\int_{0}^{\infty}\textup{d}x\,e^{-\alpha x}x^{\beta}\int_{0}^{\infty}\textup{d}y\,e^{y\left(1-x\right)}\int_{y}^{\infty}\textup{d}z\,e^{-z}z^{\gamma} &= \int_{0}^{1}\textup{d}x\,e^{-\alpha x}x^{\beta}\int_{0}^{\infty}\textup{d}y\,e^{y\left(1-x\right)}\int_{y}^{\infty}\textup{d}z\,e^{-z}z^{\gamma} + \\
 &+ \int_{1}^{\infty}\textup{d}x\,e^{-\alpha x}x^{\beta}\int_{0}^{\infty}\textup{d}y\,e^{y\left(1-x\right)}\int_{y}^{\infty}\textup{d}z\,e^{-z}z^{\gamma},
\label{triple_1}
\end{align}
and changing the order of integration through $\int_{0}^{\infty}\textup{d}y\int_{y}^{\infty}\textup{d}z\,\left(\cdots\right) = \int_{0}^{\infty}\textup{d}z\int_{0}^{z}\textup{d}y\,\left(\cdots\right)$ leads to
\begin{align}
\nonumber\lefteqn{\int_{0}^{\infty}\textup{d}x\,e^{-\alpha x}x^{\beta}\int_{0}^{\infty}\textup{d}y\,e^{y\left(1-x\right)}\int_{y}^{\infty}\textup{d}z\,e^{-z}z^{\gamma} =}& \\
\nonumber &=\int_{0}^{1}\textup{d}x\,e^{-\alpha x}x^{\beta}\int_{0}^{\infty}\textup{d}z\,e^{-z}z^{\gamma}\int_{0}^{z}\textup{d}y\,e^{y\left(1-x\right)} + \int_{1}^{\infty}\textup{d}x\,e^{-\alpha x}x^{\beta}\int_{0}^{\infty}\textup{d}z\,e^{-z}z^{\gamma}\int_{0}^{z}\textup{d}y\,e^{y\left(1-x\right)} \\
\nonumber &= \int_{0}^{1}\textup{d}x\,e^{-\alpha x}\frac{x^{\beta}}{1-x}\int_{0}^{\infty}\textup{d}z\,z^{\gamma}\left(e^{-zx} - e^{-z}\right) + \int_{1}^{\infty}\textup{d}x\,e^{-\alpha x}\frac{x^{\beta}}{1-x}\int_{0}^{\infty}\textup{d}z\,z^{\gamma}\left(e^{-zx} - e^{-z}\right) \\
\nonumber &= \int_{0}^{1}\textup{d}x\,e^{-\alpha x}\frac{x^{\beta}}{1-x}\left[\Gamma(\gamma+1)\left(\frac{1}{x^{\gamma+1}}-1\right)\right] + \int_{1}^{\infty}\textup{d}x\,e^{-\alpha x}\frac{x^{\beta}}{1-x}\left[\Gamma(\gamma+1)\left(\frac{1}{x^{\gamma+1}}-1\right)\right] \\
 &= \Gamma(\gamma+1)\sum_{n=\beta-\gamma-1}^{\beta-1}\left[ \int_{0}^{1}\textup{d}x\,e^{-\alpha x}x^{n} + \int_{1}^{\infty}\textup{d}x\,e^{-\alpha x}n^{n} \right].
\label{triple_2}
\end{align}
Then,
\begin{align}
\int_{0}^{\infty}\textup{d}x\,e^{-\alpha x}x^{\beta}\int_{0}^{\infty}\textup{d}y\,e^{y\left(1-x\right)}\int_{y}^{\infty}\textup{d}z\,e^{-z}z^{\gamma} = \Gamma(\gamma+1)\sum_{n=\beta-\gamma-1}^{\beta-1}\frac{\Gamma(n+1)}{\alpha^{n+1}} \qquad (\alpha,\beta,\gamma>0\textnormal{ and }\beta>\gamma).
\label{triple_gamma}
\end{align}


\section{The coefficient of the linear term - $L(k,k_{0})$}

In this section, the last term of \eqref{L} is analyzed. Since this term is associated with the factor $\theta(\Delta-1)$, it is assumed that $\Delta$ is a positive integer. Then,
\begin{align}
\nonumber k!e^{c}\sum_{\ell=0}^{\Delta-1}\frac{c^{\ell-k}}{\ell!}\frac{1}{k-\ell}\sum_{m=0}^{\Delta-1-\ell}\frac{\left(-c\right)^{m}}{m!} &= \frac{k!c^{-k}}{\left(\Delta-1\right)!}\int_{1}^{\infty}\textup{d}w\,w^{-k-1}\int_{-c}^{\infty}\textup{d}t\,e^{-t}\sum_{\ell=0}^{\Delta-1}{\Delta-1\choose\ell}t^{\Delta-1-\ell}\left(cw\right)^{\ell} \\
\nonumber&= \frac{k!c^{-k}}{\left(\Delta-1\right)!}\int_{1}^{\infty}\textup{d}w\,w^{-k-1}\int_{-c}^{\infty}\textup{d}t\,e^{-t}\left(t+cw\right)^{\Delta-1} \\
&= \frac{1}{\left(\Delta-1\right)!}\int_{0}^{\infty}\textup{d}u\,e^{c\left(1-u\right)}u^{k}\int_{0}^{\infty}\textup{d}x\,e^{x\left(1-u\right)}\int_{x}^{\infty}\textup{d}y\,e^{-y}y^{\Delta-1},
\label{L_last1}
\end{align}
where \eqref{1/u} and \eqref{sumgamma} (together with \eqref{incgamma}) was invoked in the first passage, while \eqref{1/ualpha} and some change of variables were used in the last passage. The resulting triple integral is evaluated using \eqref{triple_gamma}, and the result is
\begin{align}
\nonumber k!e^{c}\sum_{\ell=0}^{\Delta-1}\frac{c^{\ell-k}}{\ell!}\frac{1}{k-\ell}\sum_{m=0}^{\Delta-1-\ell}\frac{\left(-c\right)^{m}}{m!} &= \sum_{n=k_{0}}^{k-1}\left[\int_{0}^{1}\textup{d}u\,e^{c\left(1-u\right)}u^{n} + \int_{1}^{\infty}\textup{d}u\,e^{c\left(1-u\right)}u^{n}\right] \\
 &= e^{c}\sum_{n=k_{0}}^{k-1}\left[\frac{\gamma(n+1,c)}{c^{n+1}} + \frac{\Gamma(n+1,c)}{c^{n+1}}\right] \qquad (k>k_{0}).
\label{L_last2}
\end{align}

Therefore, from \eqref{Omega1-Omega1} and \eqref{L_last2}, the coefficient $L(k,k_{0})$ in \eqref{L} can be cast as
\begin{align}
L(k,k_{0}) = \left\{
\begin{array}{ccl}
\displaystyle-Me^{c}\sum_{n=k_{0}}^{k-1}\frac{\Gamma(n+1,c)}{c^{n+1}} &,& k>k_{0} \\
 & & \\
\displaystyle-Me^{c}\sum_{n=k}^{k_{0}-1}\frac{\gamma(n+1,c)}{c^{n+1}} &,& k<k_{0}
\end{array}
\right..
\label{Lfinal}
\end{align}


\section{The coefficient of the quadratic term - $Q(k,k_{0})$}

By defining
\begin{align}
Q(k,k_{0}) = MQ_{1}(k,k_{0}) + M^{2}Q_{2}(k,k_{0}),
\label{Qsplit}
\end{align}
the coeffcient of the quadratic term \eqref{Q} was splitted into two parts, where
\begin{align}
\nonumber Q_{1}(k,k_{0}) &:= c\int_{0}^{1}\textup{d}\xi\,e^{c\left(1-\xi\right)}\left(\xi^{k_{0}}-\xi^{k}\right) + \frac{e^{c}k!}{c^{k}}\theta(\Delta-1)\sum_{\ell=k_{0}+1}^{k} \frac{c^{k-\ell}}{\left(k-\ell\right)!} \sum_{m=0}^{\ell-1-k_{0}}\frac{\left(-c\right)^{m}}{m!} + \\
 &+ \frac{e^{c}k!}{c^{k}}\theta(\Delta-1)\sum_{\ell=k_{0}+1}^{k} \frac{c^{k-\ell}}{\left(k-\ell\right)!\ell} \frac{\left(-c\right)^{\ell-k_{0}}}{\left(\ell-1-k_{0}\right)!}
\label{Q1}
\end{align}
and
\begin{align}
\nonumber Q_{2}(k,k_{0}) &:= \frac{1}{2}\Omega_{2}(k_{0})-\frac{1}{2}\Omega_{2}(k)+\Omega_{1}^{2}(k)-\Omega_{1}(k_{0})\Omega_{1}(k) + \\
\nonumber &+ \frac{e^{c}k!}{c^{k}}\theta(\Delta-1)\Omega_{1}(k)\sum_{\ell=k_{0}+1}^{k} \frac{c^{k-\ell}}{\left(k-\ell\right)!\ell}\sum_{m=0}^{\ell-1-k_{0}}\frac{\left(-c\right)^{m}}{m!} - \\
\nonumber & - \frac{e^{c}k!}{c^{k}}\theta(\Delta-1)\sum_{\ell=k_{0}+1}^{k} \frac{\theta(k-\ell-1)}{\ell}\sum_{m=0}^{\ell-1-k_{0}}\frac{\left(-c\right)^{m}}{m!}\sum_{q=0}^{k-\ell-1}\frac{c^{q}}{q!}\frac{1}{k-\ell-q} + \\
\nonumber &+ \frac{e^{c}k!}{c^{k}}\theta(\Delta-1)\sum_{\ell=k_{0}+1}^{k} \frac{c^{k-\ell}}{\left(k-\ell\right)!}\frac{\theta(\ell-2-k_{0})}{\ell}\sum_{m=0}^{\ell-2-k_{0}}\frac{1}{\ell-1-k_{0}-m}\sum_{q=0}^{m}\frac{\left(-c\right)^{q}}{q!} + \\
 &+ \frac{e^{c}k!}{c^{2k}}\theta(\Delta-1)\sum_{\ell=k_{0}+1}^{k} \frac{k!}{\left(k-\ell\right)!}\frac{c^{k-\ell}}{\ell}\theta(k-1)\sum_{m=0}^{\ell-1-k_{0}}\frac{\left(-c\right)^{m}}{m!}\sum_{q=0}^{k-1}\frac{c^{q}}{q!}\frac{1}{k-q}.
\label{Q2}
\end{align}
The functions $Q_{1}$ and $Q_{2}$ are, respectively, the coefficients of the second-order terms (in $\alpha$) associated to $M$ and $M^{2}$. Naturally, the dominant term of $Q$ is given by $M^{2}Q_{2}$, but the exact expression for $Q_{1}$ will also be determined, which is useful for numerical purposes.


\subsection{The coefficient of the quadratic term - the $Q_{1}(k,k_{0})$ term}

Starting from \eqref{Q1}, invoking \eqref{expmon} and \eqref{sumgamma} to deal with the integral, and introducing the variable $u=\ell-1-k_{0}$ leads to
\begin{align}
\nonumber Q_{1}(k,k_{0}) &:= e^{c}\left[\frac{\gamma(k_{0}+1,c)}{c^{k_{0}}} - \frac{\gamma(k+1,c)}{c^{k}}\right] + \frac{e^{c}k!}{c^{k}}\theta(\Delta-1)\Bigg[ \sum_{u=0}^{\Delta-1}\frac{c^{\Delta-1-u}}{\left(\Delta-1-u\right)!}\sum_{m=0}^{u}\frac{\left(-c\right)^{m}}{m!} + \\
 &+ \sum_{u=0}^{\Delta-1}\frac{c^{\Delta-1-u}}{\left(\Delta-1-u\right)!u!}\frac{\left(-c\right)^{u+1}}{u+k_{0}+1} \Bigg].
\label{Q1_a}
\end{align}
From \eqref{sumgamma}, \eqref{incgamma} and the integral representation of the Gamma function, one has
\begin{align}
\nonumber\sum_{u=0}^{\Delta-1}\frac{c^{\Delta-1-u}}{\left(\Delta-1-u\right)!}\sum_{m=0}^{u}\frac{\left(-c\right)^{m}}{m!} &= \sum_{u=0}^{\Delta-1}\frac{c^{\Delta-1-u}}{\left(\Delta-1-u\right)!}\frac{e^{-c}}{u!}\int_{-c}^{\infty}\textup{d}t\,e^{-t}t^{u} \\
\nonumber &= \frac{e^{-c}}{\left(\Delta-1\right)!}\int_{-c}^{\infty}\textup{d}t\,e^{-t}\sum_{u=0}^{\Delta-1}{\Delta-1\choose u}c^{\Delta-1-u}t^{u} \\
\nonumber &= \frac{e^{-c}}{\left(\Delta-1\right)!}\int_{-c}^{\infty}\textup{d}t\,e^{-t}\left(t+c\right)^{\Delta-1} \\
 & = 1.
\label{Q1_b}
\end{align}
Furthermore, from \eqref{1/u}, it is also possible to show that
\begin{align}
\nonumber\sum_{u=0}^{\Delta-1}\frac{c^{\Delta-1-u}}{\left(\Delta-1-u\right)!u!}\frac{\left(-c\right)^{u+1}}{u+k_{0}+1} &= \sum_{u=0}^{\Delta-1}\frac{c^{\Delta-1-u}}{\left(\Delta-1-u\right)!u!}\left(-c\right)^{u+1}\int_{1}^{\infty}\textup{d}w\,w^{-\left(u+k_{0}+1\right)-1} \\
\nonumber &= \frac{\left(-c\right)}{\left(\Delta-1\right)!}\int_{1}^{\infty}\textup{d}w\,w^{-k-1}\sum_{u=0}^{\Delta-1}{\Delta-1\choose u}\left(cw\right)^{\Delta-1-u}\left(-c\right)^{u} \\
\nonumber &= \frac{\left(-c\right)}{\left(\Delta-1\right)!}\int_{1}^{\infty}\textup{d}w\,w^{-k-1}\left(cw-c\right)^{\Delta-1} \\
 &= -\frac{c^{\Delta}}{\left(\Delta-1\right)!}\int_{0}^{\infty}\textup{d}y\,y^{\Delta-1}\frac{1}{\left(y+1\right)^{k+1}},
\label{Q1_c}
\end{align}
where the change of variable $y=w-1$ was performed in the last passage. Invoking \eqref{1/ualpha}, one can cast \eqref{Q1_c} as
\begin{align}
\nonumber\sum_{u=0}^{\Delta-1}\frac{c^{\Delta-1-u}}{\left(\Delta-1-u\right)!u!}\frac{\left(-c\right)^{u+1}}{u+k_{0}+1} &= -\frac{c^{\Delta}}{\left(\Delta-1\right)!}\int_{0}^{\infty}\textup{d}y\,y^{\Delta-1}\frac{1}{\left(y+1\right)^{k+1}} \\
\nonumber &= -\frac{c^{\Delta}}{\left(\Delta-1\right)!}\int_{0}^{\infty}\textup{d}y\,y^{\Delta-1}\frac{1}{k!}\int_{0}^{\infty}\textup{d}t\,e^{-t\left(y+1\right)}t^{k} \\
 &= -c^{\Delta}\frac{k_{0}!}{k!},
\label{Q1_d}
\end{align}
where the integration in $y$ variable was performed before $t$ in the last passage.

Inserting \eqref{Q1_b} and \eqref{Q1_d} into \eqref{Q1_a} leads to
\begin{align}
\nonumber Q_{1}(k,k_{0}) &:= e^{c}\left[\frac{\gamma(k_{0}+1,c)}{c^{k_{0}}} - \frac{\gamma(k+1,c)}{c^{k}}\right] + e^{c}\theta(\Delta-1)\Bigg[ \frac{k!}{c^{k}} - \frac{k_{0}!}{c^{k_{0}}}\Bigg] \\
 &=\left\{
\begin{array}{ccl}
\displaystyle e^{c}\left[ \frac{\Gamma(k+1,c)}{c^{k}} - \frac{\Gamma(k_{0}+1,c)}{c^{k_{0}}} \right] &,& \Delta\geq 0 \\
 & & \\
\displaystyle e^{c}\left[ \frac{\gamma(k_{0}+1,c)}{c^{k_{0}}} - \frac{\gamma(k+1,c)}{c^{k}} \right] &,& \Delta<0
\end{array}
\right.,
\label{Q1_e}
\end{align}
obtained by \eqref{Gg} (the $\Delta\geq 0$ case).


\subsection{The coefficient of the quadratic term - the $Q_{2}(k,k_{0})$ term (Part I)}

Starting from \eqref{Q2}, the introduction of the change of variable $u=\ell-1-k_{0}$ leads to
\begin{align}
\nonumber Q_{2}(k,k_{0}) &:= \frac{1}{2}\Omega_{2}(k_{0})-\frac{1}{2}\Omega_{2}(k)+\Omega_{1}^{2}(k)-\Omega_{1}(k_{0})\Omega_{1}(k) + \\
\nonumber &+ \frac{e^{c}k!}{c^{k}}\theta(\Delta-1)\Omega_{1}(k)\sum_{u=0}^{\Delta-1} \frac{c^{\Delta-1-u}}{\left(\Delta-1-u\right)!}\frac{1}{u+k_{0}+1}\sum_{m=0}^{u}\frac{\left(-c\right)^{m}}{m!} - \\
\nonumber &- \frac{e^{c}k!}{c^{k}}\theta(\Delta-1)\sum_{u=0}^{\Delta-1} \frac{\theta(\Delta-2-u)}{u+k_{0}+1}\sum_{m=0}^{u}\frac{\left(-c\right)^{m}}{m!}\sum_{q=0}^{\Delta-2-u}\frac{c^{q}}{q!}\frac{1}{\Delta-1-u-q} + \\
\nonumber &+ \frac{e^{c}k!}{c^{k}}\theta(\Delta-1)\sum_{u=0}^{\Delta-1} \frac{c^{\Delta-1-u}}{\left(\Delta-1-u\right)!}\frac{\theta(u-1)}{u+k_{0}+1}\sum_{m=0}^{u-1}\frac{1}{u-m}\sum_{q=0}^{m}\frac{\left(-c\right)^{q}}{q!} + \\
 &+ \frac{e^{c}k!}{c^{2k}}\theta(\Delta-1)\sum_{u=0}^{\Delta-1} \frac{c^{\Delta-1-u}}{\left(\Delta-1-u\right)!}\frac{k!\theta(k-1)}{u+k_{0}+1}\sum_{m=0}^{u}\frac{\left(-c\right)^{m}}{m!}\sum_{q=0}^{k-1}\frac{c^{q}}{q!}\frac{1}{k-q}.
\label{Q2_a}
\end{align}
Since $\theta(\Delta-1)\sum_{u=0}^{\Delta-1}\theta(\Delta-2-u)\left(\cdots\right)=\theta(\Delta-2)\sum_{u=0}^{\Delta-2}\left(\cdots\right)$, $\theta(\Delta-1)\sum_{u=0}^{\Delta-1}\theta(u-1)\left(\cdots\right)=\theta(\Delta-2)\sum_{u=1}^{\Delta-1}\left(\cdots\right)$ and $\theta(\Delta-1)\theta(k-1)=\theta(\Delta-1)$, the above expression can be cast as
\begin{align}
\nonumber Q_{2}(k,k_{0}) &:= \frac{1}{2}\Omega_{2}(k_{0})-\frac{1}{2}\Omega_{2}(k)+\Omega_{1}^{2}(k)-\Omega_{1}(k_{0})\Omega_{1}(k) + \\
\nonumber &+ \frac{e^{c}k!}{c^{k}}\theta(\Delta-1)\Omega_{1}(k)\sum_{u=0}^{\Delta-1} \frac{c^{\Delta-1-u}}{\left(\Delta-1-u\right)!}\frac{1}{u+k_{0}+1}\sum_{m=0}^{u}\frac{\left(-c\right)^{m}}{m!} - \\
\nonumber &- \frac{e^{c}k!}{c^{k}}\theta(\Delta-2)\sum_{u=0}^{\Delta-2} \frac{1}{u+k_{0}+1}\sum_{m=0}^{u}\frac{\left(-c\right)^{m}}{m!}\sum_{q=0}^{\Delta-2-u}\frac{c^{q}}{q!}\frac{1}{\Delta-1-u-q} + \\
\nonumber &+ \frac{e^{c}k!}{c^{k}}\theta(\Delta-2)\sum_{u=0}^{\Delta-2} \frac{c^{\Delta-2-u}}{\left(\Delta-2-u\right)!}\frac{1}{u+k_{0}+2}\sum_{m=0}^{u}\frac{1}{u+1-m}\sum_{q=0}^{m}\frac{\left(-c\right)^{q}}{q!} + \\
 &+ \frac{e^{c}k!}{c^{2k}}\theta(\Delta-1)\sum_{u=0}^{\Delta-1} \frac{c^{\Delta-1-u}}{\left(\Delta-1-u\right)!}\frac{k!}{u+k_{0}+1}\sum_{m=0}^{u}\frac{\left(-c\right)^{m}}{m!}\sum_{q=0}^{k-1}\frac{c^{q}}{q!}\frac{1}{k-q}
\label{Q2_b}
\end{align}
after a simple change of variable in the penultimate term. From $\frac{1}{\left(u+k_{0}+2\right)\left(u+1-m\right)}=\frac{1}{m+k_{0}+1}\left(\frac{1}{u+1-m}-\frac{1}{u+k_{0}+2}\right)$, the penultimate term in \eqref{Q2_b} can be written as
\begin{align}
\nonumber\lefteqn{\frac{e^{c}k!}{c^{k}}\theta(\Delta-2)\sum_{u=0}^{\Delta-2} \frac{c^{\Delta-2-u}}{\left(\Delta-2-u\right)!}\frac{1}{u+k_{0}+2}\sum_{m=0}^{u}\frac{1}{u+1-m}\sum_{q=0}^{m}\frac{\left(-c\right)^{q}}{q!} =}& \\
\nonumber &= \frac{e^{c}k!}{c^{k}}\theta(\Delta-2)\sum_{u=0}^{\Delta-2} \frac{c^{\Delta-2-u}}{\left(\Delta-2-u\right)!}\sum_{m=0}^{u}\frac{1}{m+k_{0}+1}\left(\frac{1}{u+1-m}-\frac{1}{u+k_{0}+2}\right)\sum_{q=0}^{m}\frac{\left(-c\right)^{q}}{q!} \\
 &= \frac{e^{c}k!}{c^{k}}\theta(\Delta-2)\sum_{u=0}^{\Delta-2} \frac{c^{u}}{u!}\sum_{m=0}^{\Delta-2-u}\frac{1}{m+k_{0}+1}\left(\frac{1}{\Delta-1-u-m}-\frac{1}{k-u}\right)\sum_{q=0}^{m}\frac{\left(-c\right)^{q}}{q!},
\label{Q2_c}
\end{align}
where the change of variable $u \rightarrow \Delta-2-u$ was invoked in the last passage. Inverting the order of the summation through $\sum_{u=0}^{\Delta-2}\sum_{m=0}^{\Delta-2-u}(\cdots)=\sum_{m=0}^{\Delta-2}\sum_{u=0}^{\Delta-2-m}(\cdots)$ and relabelling then some summation indices leads to
\begin{align}
\nonumber\lefteqn{\frac{e^{c}k!}{c^{k}}\theta(\Delta-2)\sum_{u=0}^{\Delta-2} \frac{c^{\Delta-2-u}}{\left(\Delta-2-u\right)!}\frac{1}{u+k_{0}+2}\sum_{m=0}^{u}\frac{1}{u+1-m}\sum_{q=0}^{m}\frac{\left(-c\right)^{q}}{q!} =}& \\
\nonumber &= -\frac{e^{c}k!}{c^{k}}\theta(\Delta-2)\sum_{u=0}^{\Delta-2}\frac{1}{u+k_{0}+1}\sum_{q=0}^{\Delta-2-u}\frac{c^{q}}{q!}\frac{1}{k-q}\sum_{m=0}^{u}\frac{\left(-c\right)^{m}}{m!} + \\
&+ \frac{e^{c}k!}{c^{k}}\theta(\Delta-2)\sum_{u=0}^{\Delta-2}\frac{1}{u+k_{0}+1}\sum_{m=0}^{u}\frac{\left(-c\right)^{m}}{m!}\sum_{q=0}^{\Delta-2-u}\frac{c^{q}}{q!}\frac{1}{\Delta-1-u-q},
\label{Q2_d}
\end{align}
and the last term matches the third line in \eqref{Q2_b}. Therefore, from \eqref{Q2_b} and \eqref{Q2_d}, one has
\begin{align}
\nonumber Q_{2}(k,k_{0}) &:= \frac{1}{2}\Omega_{2}(k_{0})-\frac{1}{2}\Omega_{2}(k)+\Omega_{1}^{2}(k)-\Omega_{1}(k_{0})\Omega_{1}(k) - \\
\nonumber &-\frac{k!e^{c}}{c^{k}}\theta(\Delta-2)\sum_{u=0}^{\Delta-2}\frac{1}{u+k_{0}+1}\sum_{q=0}^{\Delta-2-u}\frac{c^{q}}{q!}\frac{1}{k-q}\sum_{m=0}^{u}\frac{\left(-c\right)^{m}}{m!} + \\
\nonumber &+ \frac{k!e^{c}}{c^{k}}\theta(\Delta-1)\Omega_{1}(k)\sum_{u=0}^{\Delta-1} \frac{c^{\Delta-1-u}}{\left(\Delta-1-u\right)!}\frac{1}{u+k_{0}+1}\sum_{m=0}^{u}\frac{\left(-c\right)^{m}}{m!} + \\
 &+ \frac{k!e^{c}}{c^{2k}}\theta(\Delta-1)\sum_{u=0}^{\Delta-1} \frac{c^{\Delta-1-u}}{\left(\Delta-1-u\right)!}\frac{k!}{u+k_{0}+1}\sum_{m=0}^{u}\frac{\left(-c\right)^{m}}{m!}\sum_{q=0}^{k-1}\frac{c^{q}}{q!}\frac{1}{k-q}.
\label{Q2_e}
\end{align}

Let $Q_{2A}$, $Q_{2B}$ and $Q_{2C}$ be defined by
\begin{align}
\left\{
\begin{array}{ccl}
Q_{2A}(k,k_{0}) &:=&\displaystyle \frac{k!e^{c}}{c^{k}}\sum_{u=0}^{\Delta-2}\frac{1}{u+k_{0}+1}\sum_{q=0}^{\Delta-2-u}\frac{c^{q}}{q!}\frac{1}{k-q}\sum_{m=0}^{u}\frac{\left(-c\right)^{m}}{m!} \\
Q_{2B}(k,k_{0}) &:=&\displaystyle \frac{k!e^{c}}{c^{k}}\sum_{u=0}^{\Delta-1} \frac{c^{\Delta-1-u}}{\left(\Delta-1-u\right)!}\frac{1}{u+k_{0}+1}\sum_{m=0}^{u}\frac{\left(-c\right)^{m}}{m!} \\
Q_{2C}(k,k_{0}) &:=&\displaystyle \frac{k!e^{c}}{c^{2k}}\sum_{u=0}^{\Delta-1} \frac{c^{\Delta-1-u}}{\left(\Delta-1-u\right)!}\frac{k!}{u+k_{0}+1}\sum_{m=0}^{u}\frac{\left(-c\right)^{m}}{m!}\sum_{q=0}^{k-1}\frac{c^{q}}{q!}\frac{1}{k-q}
\end{array}
\right..
\label{Q2ABC}
\end{align}
These functions are part of \eqref{Q2_e}, and they are analyzed separately. All of them are associated with a factor $\theta(\Delta-1)$ or $\theta(\Delta-2)$, which allows one to concentrate on the case $\Delta>0$ only in the next three subsections.


\subsubsection{The $Q_{2A}(k,k_{0})$ function}

From \eqref{1/u} and \eqref{sumgamma} (with \eqref{incgamma}), one can cast $Q_{2A}$ as
\begin{align}
\nonumber Q_{2A}(k,k_{0}) &= \frac{k!}{c^{k}}\sum_{u=0}^{\Delta-2}\int_{1}^{\infty}\textup{d}w\,w^{-k+\left(\Delta-2-u\right)}\int_{1}^{\infty}\textup{d}z\,z^{-k-1}\sum_{q=0}^{\Delta-2-u}\frac{\left(cz\right)^{q}}{q!}\frac{1}{u!}\int_{-c}^{\infty}\textup{d}t\,e^{-t}t^{u} \\
\nonumber &= \frac{k!c^{-k}}{\left(\Delta-2\right)!}\int_{1}^{\infty}\textup{d}w\,w^{-k}\int_{1}^{\infty}\textup{d}z\,e^{cz}z^{-k-1}\int_{cz}^{\infty}\textup{d}y\,e^{-y}\int_{-c}^{\infty}\textup{d}t\,e^{-t}\sum_{u=0}^{\Delta-2}{\Delta-2\choose u}\left(wy\right)^{\Delta-2-u}t^{u} \\
 &= \frac{k!c^{-k}}{\left(\Delta-2\right)!}\int_{1}^{\infty}\textup{d}w\,w^{-k}\int_{1}^{\infty}\textup{d}z\,e^{cz}z^{-k-1}\int_{cz}^{\infty}\textup{d}y\,e^{-y}\int_{-c}^{\infty}\textup{d}t\,e^{-t}\left(wy+t\right)^{\Delta-2}.
\label{Q2A_a}
\end{align}
Introducing the change of variables $x=z-1$, $v=y/c-1$ and $s=t/c+w$ (replacing $t$ for $s$) leads to
\begin{align}
\nonumber Q_{2A}(k,k_{0}) &= \frac{k!c^{-k_{0}}}{\left(\Delta-2\right)!}\int_{1}^{\infty}\textup{d}w\,e^{cw}w^{-k}\int_{0}^{\infty}\textup{d}x\,e^{cx}\left(1+x\right)^{-k-1}\int_{x}^{\infty}\textup{d}v\,e^{-cv}\int_{w-1}^{\infty}\textup{d}s\,e^{-cs}\left(wv+s\right)^{\Delta-2} \\
\nonumber &= \frac{k!c^{-k_{0}}}{\left(\Delta-2\right)!}\sum_{n=0}^{\Delta-2}{\Delta-2\choose n}\int_{1}^{\infty}\textup{d}w\,e^{cw}w^{-\left(k-n\right)}\int_{w-1}^{\infty}\textup{d}s\,e^{-cs}s^{\Delta-2-n}\times \\
 &\times\int_{0}^{\infty}\textup{d}x\,e^{cx}\left(1+x\right)^{-k-1}\int_{x}^{\infty}\textup{d}v\,e^{-cv}v^{n}.
\label{Q2A_b}
\end{align}
Changing the order of integration through $\int_{1}^{\infty}\textup{d}w\int_{w-1}^{\infty}\textup{d}s\,\left(\cdots\right)=\int_{0}^{\infty}\textup{d}s\int_{1}^{s+1}\textup{d}w\,\left(\cdots\right)$, and then invoking \eqref{1/ualpha} leads to
\begin{align}
\nonumber Q_{2A}(k,k_{0}) &= \frac{k!c^{-k_{0}}}{\left(\Delta-2\right)!}\sum_{n=0}^{\Delta-2}{\Delta-2\choose n}\int_{0}^{\infty}\textup{d}s\,e^{-cs}s^{\Delta-2-n}\int_{1}^{s+1}\textup{d}w\,e^{cw}\left[\frac{1}{\left(k-n-1\right)!}\int_{0}^{\infty}\textup{d}z\,e^{-zw}z^{k-n-1}\right]\times \\
\nonumber &\times\int_{0}^{\infty}\textup{d}x\,e^{cx}\left[\frac{1}{k!}\int_{0}^{\infty}\textup{d}y\,e^{-y\left(1+x\right)}y^{k}\right]\int_{x}^{\infty}\textup{d}v\,e^{-cv}v^{n} \\
\nonumber &= \frac{c^{-k_{0}}}{\left(\Delta-2\right)!}\sum_{n=0}^{\Delta-2}{\Delta-2\choose n}\frac{1}{\left(k-n-1\right)!}\int_{0}^{\infty}\textup{d}z\,z^{k-n-1}\int_{0}^{\infty}\textup{d}s\,e^{-cs}s^{\Delta-2-n}\int_{1}^{s+1}\textup{d}w\,e^{\left(c-z\right)w}\times \\
 &\times\int_{0}^{\infty}\textup{d}y\,e^{-y}y^{k}\int_{0}^{\infty}\textup{d}x\,e^{x\left(c-y\right)}\int_{x}^{\infty}\textup{d}v\,e^{-cv}v^{n}.
\label{Q2A_c}
\end{align}
In \eqref{Q2A_c}, there are two independent triple integrals. The first one, which involves integration in $z$, $s$, and $w$ variables, can be performed by a similar argument used to derive \eqref{triple_gamma} to circumvent an \textquotedblleft apparent singularity\textquotedblright. The evaluation of the triple integral, which involves the variables $y$, $x$, and $v$ can be done using \eqref{triple_gamma} after some simple change of variables. After these operations and some simplifications, one has
\begin{align}
Q_{2A}(k,k_{0}) = e^{c}\sum_{n=0}^{\Delta-2}\frac{c^{k-1-n}}{\left(k-1-n\right)!}\sum_{\ell=k_{0}}^{k-n-2}\frac{\ell!}{c^{\ell+1}}\sum_{m=k-1-n}^{k-1}\frac{m!}{c^{m+1}}.
\label{Q2A_d}
\end{align}
The introduction of the change of variable $k-1-n\rightarrow n$ leads to
\begin{align}
Q_{2A}(k,k_{0}) = e^{c}\sum_{n=k_{0}+1}^{k-1}\frac{c^{n}}{n!}\sum_{\ell=k_{0}}^{n-1}\frac{\ell!}{c^{\ell+1}}\sum_{m=n}^{k-1}\frac{m!}{c^{m+1}}.
\label{Q2A_end}
\end{align}


\subsubsection{The $Q_{2B}(k,k_{0})$ function}

From \eqref{1/u} and \eqref{sumgamma} (with \eqref{incgamma}), one can cast $Q_{2B}$ as
\begin{align}
\nonumber Q_{2B}(k,k_{0}) &= \frac{k!c^{-k}}{\left(\Delta-1\right)!}\int_{1}^{\infty}\textup{d}w\,w^{-k-1}\int_{-c}^{\infty}\textup{d}t\,e^{-t}\sum_{u=0}^{\Delta-1}{\Delta-1\choose u}\left(cw\right)^{\Delta-1-u}t^{u} \\
 &= \frac{c^{-k}}{\left(\Delta-1\right)!}\int_{1}^{\infty}\textup{d}w\,\int_{0}^{\infty}\textup{d}z\,e^{-zw}z^{k}\int_{-c}^{\infty}\textup{d}t\,e^{-t}\left(cw+t\right)^{\Delta-1},
\label{Q2B_a}
\end{align}
where \eqref{1/ualpha} was invoked in the last passage. By the change of variables $x=c\left(w-1\right)$, $y=t+c$ and $r=z/c$, one has
\begin{align}
\nonumber Q_{2B}(k,k_{0}) &= \frac{e^{c}}{\left(\Delta-1\right)!}\int_{0}^{\infty}\textup{d}r\,e^{-cr}r^{k}\int_{0}^{\infty}\textup{d}x\,e^{-rx}\int_{0}^{\infty}\textup{d}y\,e^{-y}\left(x+y\right)^{\Delta-1} \\
 &= \frac{e^{c}}{\left(\Delta-1\right)!}\int_{0}^{\infty}\textup{d}r\,e^{-cr}r^{k}\int_{0}^{\infty}\textup{d}x\,e^{x\left(1-r\right)}\int_{x}^{\infty}\textup{d}v\,e^{-v}v^{\Delta-1},
\label{Q2B_b}
\end{align}
where the change of variable $v=x+y$ (changing from $y$ to $v$) was introduced. The resulting triple integral can be evaluated using \eqref{triple_gamma}, which leads to
\begin{align}
Q_{2B}(k,k_{0}) = e^{c}\sum_{n=k_{0}}^{k-1}\frac{n!}{c^{n+1}}.
\label{Q2B_end}
\end{align}


\subsubsection{The $Q_{2C}(k,k_{0})$ function}

This function can be seen as a product of two independent parts, which are
\begin{align}
\frac{e^{c}k!}{c^{k}}\sum_{u=0}^{\Delta-1} \frac{c^{\Delta-1-u}}{\left(\Delta-1-u\right)!}\frac{1}{u+k_{0}+1}\sum_{m=0}^{u}\frac{\left(-c\right)^{m}}{m!} \quad\textnormal{ and }\quad \frac{k!}{c^{k}}\sum_{q=0}^{k-1}\frac{c^{q}}{q!}\frac{1}{k-q}.
\label{Q2C_a}
\end{align}
The first one is equal to $Q_{2B}$ and, according to \eqref{Q2B_end}, can be cast as
\begin{align}
\frac{e^{c}k!}{c^{k}}\sum_{u=0}^{\Delta-1} \frac{c^{\Delta-1-u}}{\left(\Delta-1-u\right)!}\frac{1}{u+k_{0}+1}\sum_{m=0}^{u}\frac{\left(-c\right)^{m}}{m!} = e^{c}\sum_{n=k_{0}}^{k-1}\frac{n!}{c^{n+1}}.
\label{Q2C_b}
\end{align}
The second part of $Q_{2C}$, on the other hand, can be written as
\begin{align}
\nonumber \frac{k!}{c^{k}}\sum_{q=0}^{k-1}\frac{c^{q}}{q!}\frac{1}{k-q} &= \frac{k!}{c^{k}}\int_{1}^{\infty}\textup{d}w\,w^{-k-1}\sum_{q=0}^{k-1}\frac{\left(cw\right)^{q}}{q!} \\
\nonumber &= \frac{k!}{c^{k}}\int_{1}^{\infty}\textup{d}w\,\left[\frac{1}{k!}\int_{0}^{\infty}\textup{d}z\,e^{-zw}z^{k}\right]\frac{e^{cw}}{\left(k-1\right)!}\int_{cw}^{\infty}\textup{d}t\,e^{-t}t^{k-1} \\
 &= \frac{c^{-k}}{\left(k-1\right)!}\int_{1}^{\infty}\textup{d}w\,e^{cw}\int_{0}^{\infty}\textup{d}z\,e^{-zw}z^{k}\int_{cw}^{\infty}\textup{d}t\,e^{-t}t^{k-1},
\label{Q2C_c}
\end{align}
where \eqref{1/u}, \eqref{sumgamma} (and \eqref{incgamma}) and \eqref{1/ualpha} were used. Introducing the change of variables $x=c\left(w-1\right)$, $y=t-c$ and $r=z/c$ leads to
\begin{align}
\nonumber \frac{k!}{c^{k}}\sum_{q=0}^{k-1}\frac{c^{q}}{q!}\frac{1}{k-q} &= \frac{1}{\left(k-1\right)!}\int_{0}^{\infty}\textup{d}r\,e^{-cr}r^{k}\int_{0}^{\infty}\textup{d}x\,e^{x\left(1-r\right)}\int_{x}^{\infty}\textup{d}y\,e^{-y}\left(y+c\right)^{k-1} \\
\nonumber &= \frac{1}{\left(k-1\right)!}\sum_{n=0}^{k-1}{k-1\choose n}c^{k-1-n}\int_{0}^{\infty}\textup{d}r\,e^{-cr}r^{k}\int_{0}^{\infty}\textup{d}x\,e^{x\left(1-r\right)}\int_{x}^{\infty}\textup{d}y\,e^{-y}y^{n} \\
 &= \sum_{m=0}^{k-1}\frac{c^{m}}{m!}\sum_{\ell=m}^{k-1}\frac{\ell!}{c^{\ell+1}},
\label{Q2C_d}
\end{align}
where \eqref{triple_gamma} was invoked in the last passage and some simple change of variables were performed. Inverting the order of the summation in \eqref{Q2C_d} through $\sum_{m=0}^{k-1}\sum_{\ell=m}^{k-1}\left(\cdots\right)=\sum_{\ell=0}^{k-1}\sum_{m=0}^{\ell}\left(\cdots\right)$ leads to
\begin{align}
\frac{k!}{c^{k}}\sum_{q=0}^{k-1}\frac{c^{q}}{q!}\frac{1}{k-q} = \sum_{\ell=0}^{k-1}\frac{\ell!}{c^{\ell+1}}\sum_{m=0}^{\ell}\frac{c^{m}}{m!} = e^{c}\sum_{\ell=0}^{k-1}\frac{\Gamma(\ell+1,c)}{c^{\ell+1}},
\label{Q2C_e}
\end{align}
where \eqref{sumgamma} was used.

From \eqref{Q2C_b} and \eqref{Q2C_e}, one has
\begin{align}
Q_{2C}(k,k_{0}) = \left[e^{c}\sum_{n=k_{0}}^{k-1}\frac{n!}{c^{n+1}}\right]\left[e^{c}\sum_{\ell=0}^{k-1}\frac{\Gamma(\ell+1,c)}{c^{\ell+1}}\right].
\label{Q2C_end}
\end{align}


\subsection{The coefficient of the quadratic term - the $Q_{2}(k,k_{0})$ term (Part II)}

After inserting \eqref{Q2A_end}, \eqref{Q2B_end} and \eqref{Q2C_end} into \eqref{Q2_e}, one has
\begin{align}
\nonumber Q_{2}(k,k_{0}) &= \frac{1}{2}\Omega_{2}(k_{0})-\frac{1}{2}\Omega_{2}(k)+\Omega_{1}(k)\left[\Omega_{1}(k)-\Omega_{1}(k_{0})\right] -\theta(\Delta-2)e^{c}\sum_{n=k_{0}+1}^{k-1}\frac{c^{n}}{n!}\sum_{\ell=k_{0}}^{n-1}\frac{\ell!}{c^{\ell+1}}\sum_{m=n}^{k-1}\frac{m!}{c^{m+1}} + \\
 &+ \theta(\Delta-1)\Omega_{1}(k)e^{c}\sum_{n=k_{0}}^{k-1}\frac{n!}{c^{n+1}} + \theta(\Delta-1)\left[e^{c}\sum_{n=k_{0}}^{k-1}\frac{n!}{c^{n+1}}\right]\left[e^{c}\sum_{\ell=0}^{k-1}\frac{\Gamma(\ell+1,c)}{c^{\ell+1}}\right].
\label{Q2_f}
\end{align}
Note that
\begin{align}
\nonumber \lefteqn{-\theta(\Delta-2)e^{c}\sum_{n=k_{0}+1}^{k-1}\frac{c^{n}}{n!}\sum_{\ell=k_{0}}^{n-1}\frac{\ell!}{c^{\ell+1}}\sum_{m=n}^{k-1}\frac{m!}{c^{m+1}} =}& \\
\nonumber &= -\theta(\Delta-2)e^{c}\sum_{n=k_{0}+1}^{k-1}\frac{c^{n}}{n!}\left[\sum_{\ell=k_{0}}^{n}\frac{\ell!}{c^{\ell+1}} - \frac{n!}{c^{n+1}}\right]\sum_{m=n}^{k-1}\frac{m!}{c^{m+1}} \\
 &= \theta(\Delta-2)e^{c}\left[ \frac{1}{c}\sum_{n=k_{0}+1}^{k-1}\sum_{m=n}^{k-1}\frac{m!}{c^{m+1}} - \sum_{n=k_{0}+1}^{k-1}\frac{c^{n}}{n!}\sum_{\ell=k_{0}}^{n}\frac{\ell!}{c^{\ell+1}}\sum_{m=n}^{k-1}\frac{m!}{c^{m+1}} \right].
\label{D-2term_a}
\end{align}
Introducing the change in the summation order $\sum_{n=k_{0}+1}^{k-1}\sum_{m=n}^{k-1}\left(\cdots\right)=\sum_{m=k_{0}+1}^{k-1}\sum_{n=k_{0}+1}^{m}\left(\cdots\right)$ in the first sum, and splitting the second sum leads to
\begin{align}
\nonumber \lefteqn{-\theta(\Delta-2)e^{c}\sum_{n=k_{0}+1}^{k-1}\frac{c^{n}}{n!}\sum_{\ell=k_{0}}^{n-1}\frac{\ell!}{c^{\ell+1}}\sum_{m=n}^{k-1}\frac{m!}{c^{m+1}} =}& \\
\nonumber &= \theta(\Delta-2)e^{c}\left\{ \frac{1}{c}\sum_{m=k_{0}+1}^{k-1}\frac{m!}{c^{m+1}}\sum_{n=k_{0}+1}^{m}1 - \left[\sum_{n=k_{0}}^{k-1}\frac{c^{n}}{n!}\sum_{\ell=k_{0}}^{n}\frac{\ell!}{c^{\ell+1}}\sum_{m=n}^{k-1}\frac{m!}{c^{m+1}} - \frac{c^{k_{0}}}{k_{0}!}\sum_{\ell=k_{0}}^{k_{0}}\frac{\ell!}{c^{\ell+1}}\sum_{m=k_{0}}^{k-1}\frac{m!}{c^{m+1}}\right] \right\} \\
 &= \theta(\Delta-1)e^{c}\left[ \frac{1}{c}\sum_{m=k_{0}}^{k-1}\frac{m!}{c^{m+1}}\left(m-k_{0}+1\right) - \sum_{n=k_{0}}^{k-1}\frac{c^{n}}{n!}\sum_{\ell=k_{0}}^{n}\frac{\ell!}{c^{\ell+1}}\sum_{m=n}^{k-1}\frac{m!}{c^{m+1}} \right].
\label{D-2term_b}
\end{align}
In the last passage, since the expression inside the square brackets in \eqref{D-2term_b} (last line) vanishes when $\Delta=1$ (or $k=k_{0}+1$), one can replace the the unit step function $\theta(\Delta-2)$ by $\theta(\Delta-1)$ without causing any impact on the whole term. Then, changing the order of the summation through $\sum_{n=k_{0}}^{k-1}\sum_{\ell=k_{0}}^{n}\sum_{m=n}^{k-1}\left(\cdots\right)=\sum_{\ell=k_{0}}^{k-1}\sum_{m=\ell}^{k-1}\sum_{n=\ell}^{m}\left(\cdots\right)$ implies
\begin{align}
\nonumber \lefteqn{-\theta(\Delta-2)e^{c}\sum_{n=k_{0}+1}^{k-1}\frac{c^{n}}{n!}\sum_{\ell=k_{0}}^{n-1}\frac{\ell!}{c^{\ell+1}}\sum_{m=n}^{k-1}\frac{m!}{c^{m+1}} =}& \\
\nonumber &= \theta(\Delta-1)e^{c}\left[ \frac{1}{c}\sum_{m=k_{0}}^{k-1}\frac{m!}{c^{m+1}}\left(m-k_{0}+1\right) - \sum_{\ell=k_{0}}^{k-1}\frac{\ell!}{c^{\ell+1}}\sum_{m=\ell}^{k-1}\frac{m!}{c^{m+1}}\sum_{n=\ell}^{m}\frac{c^{n}}{n!} \right] \\
\nonumber &= \theta(\Delta-1)e^{c}\left[ \frac{1}{c}\sum_{m=k_{0}}^{k-1}\frac{m!}{c^{m+1}}\left(m-k_{0}+1\right) - \sum_{\ell=k_{0}}^{k-1}\frac{\ell!}{c^{\ell+1}}\sum_{m=\ell}^{k-1}\frac{m!}{c^{m+1}}\left(\sum_{n=0}^{m}\frac{c^{n}}{n!} - \sum_{n=0}^{\ell}\frac{c^{n}}{n!} + \frac{c^{\ell}}{\ell!}\right) \right] \\
 &= \theta(\Delta-1)e^{2c}\left[ \sum_{\ell=k_{0}}^{k-1}\frac{\Gamma(\ell+1,c)}{c^{\ell+1}}\sum_{m=\ell}^{k-1}\frac{m!}{c^{m+1}} - \sum_{\ell=k_{0}}^{k-1}\frac{\ell!}{c^{\ell+1}}\sum_{m=\ell}^{k-1}\frac{\Gamma(m+1,c)}{c^{m+1}} \right],
\label{D-2term_c}
\end{align}
where \eqref{sumgamma} and \eqref{recurrencegamma}, and a change in summation order $\sum_{\ell=k_{0}}^{k-1}\sum_{m=\ell}^{k-1}\left(\cdots\right)=\sum_{m=k_{0}}^{k-1}\sum_{\ell=k_{0}}^{m}\left(\cdots\right)$ were used in the last passage. Inserting \eqref{D-2term_c} back to \eqref{Q2_f} yields
\begin{align}
\nonumber Q_{2}(k,k_{0}) &= \frac{1}{2}\Big[\Omega_{2}(k_{0})-\Omega_{2}(k)\Big]+\Omega_{1}(k)\Big[\Omega_{1}(k)-\Omega_{1}(k_{0})\Big] + \theta(\Delta-1)\Omega_{1}(k)e^{c}\sum_{n=k_{0}}^{k-1}\frac{n!}{c^{n+1}} + \\
 & + \theta(\Delta-1)e^{2c}\sum_{n=k_{0}}^{k-1}\frac{n!}{c^{n+1}}\sum_{\ell=k_{0}}^{n}\frac{\Gamma(\ell+1,c)}{c^{\ell+1}} + \theta(\Delta-1)e^{2c}\sum_{n=k_{0}}^{k-1}\frac{n!}{c^{n+1}}\sum_{\ell=0}^{n-1}\frac{\Gamma(\ell+1,c)}{c^{\ell+1}}
\label{Q2_end}
\end{align}
after changing the order of one of sums through $\sum_{\ell=k_{0}}^{k-1}\sum_{n=\ell}^{k-1}\left(\cdots\right)=\sum_{n=k_{0}}^{k-1}\sum_{\ell=k_{0}}^{n}\left(\cdots\right)$ and some trivial relabelling of the indices.

From now on, it is convenient to divide the analysis into the $\Delta>0$ and $\Delta<0$ cases.


\subsection{The coefficient of the quadratic term - the $Q_{2}(k,k_{0})$ term (case $\Delta>0$)}

Define by $Q_{>}$ the coefficient $Q_{2}$ when $k>k_{0}$. From \eqref{Omega2-Omega2}, \eqref{Omega1-Omega1} and \eqref{Q2_end} when $\Delta>0$, one has
\begin{align}
\nonumber Q_{>}(k,k_{0}) &= \sum_{n=k_{0}}^{k-1}\psi(n) + \Omega_{1}(k)e^{c}\sum_{n=k_{0}}^{k-1}\frac{\Gamma(n+1,c)}{c^{n+1}}  + e^{2c}\sum_{n=k_{0}}^{k-1}\frac{n!}{c^{n+1}}\sum_{\ell=k_{0}}^{n}\frac{\Gamma(\ell+1,c)}{c^{\ell+1}} + \\
 &+ e^{2c}\sum_{n=k_{0}}^{k-1}\frac{n!}{c^{n+1}}\sum_{\ell=0}^{n-1}\frac{\Gamma(\ell+1,c)}{c^{\ell+1}}
\label{Q>}
\end{align}
after some simplifications. From \eqref{psiOmega1} and \eqref{Omega1-Omega1}, one has
\begin{align}
\nonumber\sum_{n=k_{0}}^{k-1}\psi(n) &= \sum_{n=k_{0}}^{k-1}\left[-\frac{n!}{c^{n+1}}\sum_{\ell=0}^{n}\frac{c^{\ell}}{\ell!}\Omega_{1}(\ell)\right] \\
\nonumber &= \sum_{n=k_{0}}^{k-1}\left\{-\frac{n!}{c^{n+1}}\sum_{\ell=0}^{n}\frac{c^{\ell}}{\ell!}\left[\Omega_{1}(k) + e^{c}\sum_{m=\ell}^{k-1}\frac{\gamma(m+1,c)}{c^{m+1}}\right]\right\} \\
\nonumber &= -\Omega_{1}(k)\sum_{n=k_{0}}^{k-1}\frac{n!}{c^{n+1}}\sum_{\ell=0}^{n}\frac{c^{\ell}}{\ell!} - e^{c}\sum_{n=k_{0}}^{k-1}\frac{n!}{c^{n+1}}\sum_{\ell=0}^{n}\frac{c^{\ell}}{\ell!}\sum_{m=\ell}^{k-1}\frac{\left[m!-\Gamma(m+1,c)\right]}{c^{m+1}} \\
 &= -\Omega_{1}(k)\sum_{n=k_{0}}^{k-1}\frac{n!}{c^{n+1}}\sum_{\ell=0}^{n}\frac{c^{\ell}}{\ell!} + e^{c}\sum_{n=k_{0}}^{k-1}\frac{n!}{c^{n+1}}\sum_{\ell=0}^{n}\frac{c^{\ell}}{\ell!}\sum_{m=\ell}^{k-1}\frac{\Gamma(m+1,c)}{c^{m+1}} - e^{c}\sum_{n=k_{0}}^{k-1}\frac{n!}{c^{n+1}}\sum_{\ell=0}^{n}\frac{c^{\ell}}{\ell!}\sum_{m=\ell}^{k-1}\frac{m!}{c^{m+1}}.
\label{sumpsi>_a}
\end{align}
Changing the order of the sum through $\sum_{\ell=0}^{n}\sum_{m=\ell}^{k-1}\left(\cdots\right) = \sum_{m=0}^{n-1}\sum_{\ell=0}^{m}\left(\cdots\right) + \sum_{m=n}^{k-1}\sum_{\ell=0}^{n}\left(\cdots\right)$ in the last term (here, $n<k$), and using \eqref{sumgamma} leads to
\begin{align}
\nonumber\sum_{n=k_{0}}^{k-1}\psi(n) &= -\Omega_{1}(k)\sum_{n=k_{0}}^{k-1}\frac{e^{c}\Gamma(n+1,c)}{c^{n+1}} + e^{c}\sum_{n=k_{0}}^{k-1}\frac{n!}{c^{n+1}}\sum_{\ell=0}^{n}\frac{c^{\ell}}{\ell!}\sum_{m=\ell}^{k-1}\frac{\Gamma(m+1,c)}{c^{m+1}} - \\
\nonumber &- e^{c}\sum_{n=k_{0}}^{k-1}\frac{n!}{c^{n+1}}\sum_{m=0}^{n-1}\frac{e^{c}\Gamma(m+1,c)}{c^{m+1}} - e^{c}\sum_{n=k_{0}}^{k-1}\frac{e^{c}\Gamma(n+1,c)}{c^{n+1}}\sum_{m=n}^{k-1}\frac{m!}{c^{m+1}} \\
\nonumber&= -\Omega_{1}(k)e^{c}\sum_{n=k_{0}}^{k-1}\frac{\Gamma(n+1,c)}{c^{n+1}} + e^{c}\sum_{n=k_{0}}^{k-1}\frac{n!}{c^{n+1}}\sum_{\ell=0}^{n}\frac{c^{\ell}}{\ell!}\sum_{m=\ell}^{k-1}\frac{\Gamma(m+1,c)}{c^{m+1}} - \\
 &- e^{2c}\sum_{n=k_{0}}^{k-1}\frac{n!}{c^{n+1}}\sum_{m=0}^{n-1}\frac{\Gamma(m+1,c)}{c^{m+1}} - e^{2c}\sum_{n=k_{0}}^{k-1}\frac{n!}{c^{n+1}}\sum_{m=k_{0}}^{n}\frac{\Gamma(m+1,c)}{c^{m+1}},
\label{sumpsi>_b}
\end{align}
where the change of summation order $\sum_{n=k_{0}}^{k-1}\sum_{m=n}^{k-1}\left(\cdots\right)=\sum_{m=k_{0}}^{k-1}\sum_{n=k_{0}}^{m}\left(\cdots\right)$ (followed by a relabelling of the indices $n\leftrightarrow m$) was performed in the last term in the passage. Inserting \eqref{sumpsi>_b} back to \eqref{Q>} implies
\begin{align}
Q_{>}(k,k_{0}) = e^{c}\sum_{n=k_{0}}^{k-1}\frac{n!}{c^{n+1}}\sum_{\ell=0}^{n}\frac{c^{\ell}}{\ell!}\sum_{m=\ell}^{k-1}\frac{\Gamma(m+1,c)}{c^{m+1}}.
\label{Q>_shortest}
\end{align}
After changing the order of the sum through $\sum_{\ell=0}^{n}\sum_{m=\ell}^{k-1}\left(\cdots\right) = \sum_{m=0}^{n-1}\sum_{\ell=0}^{m}\left(\cdots\right)+\sum_{m=n}^{k-1}\sum_{\ell=0}^{n}\left(\cdots\right)$ (here, $n\leq k-1$) and invoking \eqref{sumgamma}, one has
\begin{align}
Q_{>}(k,k_{0}) = e^{k_{0}}\sum_{n=k_{0}}^{k-1}\frac{n!}{k_{0}^{n+1}}\sum_{m=0}^{n-1}\frac{e^{k_{0}}\Gamma^{2}(m+1,k_{0})}{m!k_{0}^{m+1}} + e^{k_{0}}\sum_{n=k_{0}}^{k-1}\frac{e^{k_{0}}\Gamma(n+1,k_{0})}{k_{0}^{n+1}}\sum_{m=n}^{k-1}\frac{\Gamma(m+1,k_{0})}{k_{0}^{m+1}}.
\label{Q>_c}
\end{align}
The coefficient $Q_{>}(k,k_{0})$ admits an alternative representation. Noting that
\begin{align}
\nonumber\sum_{n=k_{0}}^{k-1}\frac{\Gamma(n+1,c)}{c^{n+1}}\sum_{m=n}^{k-1}\frac{\Gamma(m+1,c)}{c^{m+1}} &= \sum_{n=k_{0}}^{k-1}\frac{\Gamma(n+1,c)}{c^{n+1}}\sum_{m=k_{0}}^{n}\frac{\Gamma(m+1,c)}{c^{m+1}} \\
 &= \frac{1}{2}\left[\sum_{n=k_{0}}^{k-1}\frac{\Gamma(n+1,c)}{c^{n+1}}\right]^{2} + \frac{1}{2}\sum_{n=k_{0}}^{k-1}\left[\frac{\Gamma(n+1,c)}{c^{n+1}}\right]^{2},
\label{Q>_aux}
\end{align}
the $Q_{>}$ function can be cast as
\begin{align}
Q_{>}(k,k_{0}) = \frac{1}{2}\left[\sum_{n=k_{0}}^{k-1}\frac{e^{c}\Gamma(n+1,c)}{c^{n+1}}\right]^{2} - \frac{1}{2}\sum_{n=k_{0}}^{k-1}\left[\frac{e^{c}\Gamma(n+1,c)}{c^{n+1}}\right]^{2} + \sum_{n=k_{0}}^{k-1}\frac{n!}{c^{n+1}}\sum_{m=0}^{n}\frac{c^{m+1}}{m!}\left[\frac{e^{c}\Gamma(m+1,c)}{c^{m+1}}\right]^{2}
\label{Q>_end}
\end{align}
after some simple manipulations.


\subsection{The coefficient of the quadratic term - the $Q_{2}(k,k_{0})$ term (case $\Delta<0$)}

Define by $Q_{<}$ the coefficient $Q_{2}$ when $k<k_{0}$. From \eqref{psiOmega1}, \eqref{Omega1-Omega1} and \eqref{Omega2-Omega2} into \eqref{Q2_end} when $\Delta<0$, one has
\begin{align}
\nonumber Q_{<}(k,k_{0}) &= -\sum_{n=k}^{k_{0}-1}\psi(n) + \Omega_{1}(k)e^{c}\sum_{n=k}^{k_{0}-1}\frac{\gamma(n+1,c)}{c^{n+1}} \\
 &= \sum_{n=k}^{k_{0}-1}\frac{n!}{c^{n+1}}\sum_{\ell=0}^{n}\frac{c^{\ell}}{\ell!}\Omega_{1}(\ell) + \Omega_{1}(k)e^{c}\sum_{n=k}^{k_{0}-1}\frac{\gamma(n+1,c)}{c^{n+1}}.
\label{Q<}
\end{align}
Since from \eqref{sumgamma}, \eqref{sumOmega1=0} and \eqref{Omega1-Omega1} the relation
\begin{align}
\nonumber\sum_{\ell=0}^{n}\frac{c^{\ell}}{\ell!}\Omega_{1}(\ell) &= \sum_{\ell=0}^{\infty}\frac{c^{\ell}}{\ell!}\Omega_{1}(\ell) - \sum_{\ell=n+1}^{\infty}\frac{c^{\ell}}{\ell!}\Omega_{1}(\ell) \\
\nonumber &= 0 - \sum_{\ell=n+1}^{\infty}\frac{c^{\ell}}{\ell!}\Omega_{1}(\ell) \\
\nonumber &= - \sum_{\ell=n+1}^{\infty}\frac{c^{\ell}}{\ell!}\left[\Omega_{1}(k) - e^{c}\sum_{m=k}^{\ell-1}\frac{\gamma(m+1,c)}{c^{m+1}}\right] \\
&= - \Omega_{1}(k)e^{c}\frac{\gamma(n+1,c)}{n!} + e^{c}\sum_{\ell=n+1}^{\infty}\frac{c^{\ell}}{\ell!}\sum_{m=k}^{\ell-1}\frac{\gamma(m+1,c)}{c^{m+1}}
\label{Q<_int}
\end{align}
can be established, one has
\begin{align}
Q_{<}(k,k_{0}) = e^{c}\sum_{n=k}^{k_{0}-1}\frac{n!}{c^{n+1}}\sum_{\ell=n+1}^{\infty}\frac{c^{\ell}}{\ell!}\sum_{m=k}^{\ell-1}\frac{\gamma(m+1,c)}{c^{m+1}}.
\label{Q<_shortest}
\end{align}
Changing the order of the sums through $\sum_{\ell=n+1}^{\infty}\sum_{m=k}^{\ell-1}\left(\cdots\right)=\sum_{m=k}^{n}\sum_{\ell=n+1}^{\infty}\left(\cdots\right)+\sum_{m=n+1}^{\infty}\sum_{\ell=m+1}^{\infty}\left(\cdots\right)$ and invoking \eqref{sumgamma} leads to
\begin{align}
Q_{<}(k,k_{0}) = e^{2c}\sum_{n=k}^{k_{0}-1}\frac{\gamma(n+1,c)}{c^{n+1}}\sum_{m=k}^{n}\frac{\gamma(m+1,c)}{c^{m+1}} + e^{2c}\sum_{n=k}^{k_{0}-1}\frac{n!}{c^{n+1}}\sum_{m=n+1}^{\infty}\frac{\gamma^{2}(m+1,c)}{m!c^{m+1}}.
\label{Q<_a}
\end{align}
Then, following a procedure similar that transformed \eqref{Q>_c} to \eqref{Q>_end}, one can finally cast
\begin{align}
Q_{<}(k,k_{0}) = \frac{1}{2}\left[\sum_{n=k}^{k_{0}-1}\frac{e^{c}\gamma(n+1,c)}{c^{n+1}}\right]^{2} - \frac{1}{2}\sum_{n=k}^{k_{0}-1}\left[\frac{e^{c}\gamma(n+1,c)}{c^{n+1}}\right]^{2} + \sum_{n=k}^{k_{0}-1}\frac{n!}{c^{n+1}}\sum_{m=n}^{\infty}\frac{c^{m+1}}{m!}\left[\frac{e^{c}\gamma(m+1,c)}{c^{m+1}}\right]^{2}.
\label{Q<_end}
\end{align}


\section{Summary}

In the expansion of the first-passage function
\begin{align}
f_{s}^{z}(k|k_{0};z) = 1 + \alpha L(k,k_{0}) + \alpha^{2}Q(k,k_{0}) + \mathcal{O}(\alpha^{3}),
\label{Fz_end}
\end{align}
one has
\begin{align}
L(k,k_{0}) = \left\{
\begin{array}{ccl}
\displaystyle-Me^{c}\sum_{n=k_{0}}^{k-1}\frac{\Gamma(n+1,c)}{c^{n+1}} &,& k>k_{0} \\
 & & \\
\displaystyle-Me^{c}\sum_{n=k}^{k_{0}-1}\frac{\gamma(n+1,c)}{c^{n+1}} &,& k<k_{0}
\end{array}
\right..
\label{Lend}
\end{align}
and
\begin{align}
Q(k,k_{0}) = MQ_{1}(k,k_{0}) + M^{2}Q_{2}(k,k_{0}),
\label{Q1Q2}
\end{align}
where
\begin{align}
Q_{1}(k,k_{0}) =\left\{
\begin{array}{ccl}
\displaystyle e^{c}\left[ \frac{\Gamma(k+1,c)}{c^{k}} - \frac{\Gamma(k_{0}+1,c)}{c^{k_{0}}} \right] &,& k>k_{0} \\
 & & \\
\displaystyle e^{c}\left[ \frac{\gamma(k_{0}+1,c)}{c^{k_{0}}} - \frac{\gamma(k+1,c)}{c^{k}} \right] &,& k<k_{0}
\end{array}
\right.
\label{Q1_end}
\end{align}
and
\begin{align}
Q_{2}(k,k_{0}) = \left\{
\begin{array}{ccl}
\displaystyle e^{c}\sum_{n=k_{0}}^{k-1}\frac{n!}{c^{n+1}}\sum_{\ell=0}^{n}\frac{c^{\ell}}{\ell!}\sum_{m=\ell}^{k-1}\frac{\Gamma(m+1,c)}{c^{m+1}} &,& k>k_{0} \\
 & & \\
\displaystyle e^{c}\sum_{n=k}^{k_{0}-1}\frac{n!}{c^{n+1}}\sum_{\ell=n+1}^{\infty}\frac{c^{\ell}}{\ell!}\sum_{m=k}^{\ell-1}\frac{\gamma(m+1,c)}{c^{m+1}} &,& k<k_{0}
\end{array}
\right.
\label{Q2_shortest}
\end{align}
or
\begin{align}
\nonumber\lefteqn{Q_{2}(k,k_{0}) =}&\\
 &=\left\{
\begin{array}{ccl}
\displaystyle \frac{1}{2}\left[\sum_{n=k_{0}}^{k-1}\frac{e^{c}\Gamma(n+1,c)}{c^{n+1}}\right]^{2} - \frac{1}{2}\sum_{n=k_{0}}^{k-1}\left[\frac{e^{c}\Gamma(n+1,c)}{c^{n+1}}\right]^{2} + \sum_{n=k_{0}}^{k-1}\frac{n!}{c^{n+1}}\sum_{m=0}^{n}\frac{c^{m+1}}{m!}\left[\frac{e^{c}\Gamma(m+1,c)}{c^{m+1}}\right]^{2} &,& k>k_{0} \\
 & & \\
\displaystyle \frac{1}{2}\left[\sum_{n=k}^{k_{0}-1}\frac{e^{c}\gamma(n+1,c)}{c^{n+1}}\right]^{2} - \frac{1}{2}\sum_{n=k}^{k_{0}-1}\left[\frac{e^{c}\gamma(n+1,c)}{c^{n+1}}\right]^{2} + \sum_{n=k}^{k_{0}-1}\frac{n!}{c^{n+1}}\sum_{m=n}^{\infty}\frac{c^{m+1}}{m!}\left[\frac{e^{c}\gamma(m+1,c)}{c^{m+1}}\right]^{2} &,& k<k_{0}
\end{array}
\right..
\label{Q2_end}
\end{align}


\begin{thebibliography}{999}

\bibitem{AB02} R. Albert and A. -L. Barab\'asi, Rev. Mod. Phys. {\bf 74}, 47 (2002).

\bibitem{DM03} S. N. Dorogovtsev and J. F. F. Mendes, \textit{Evolution of Networks} (Oxford University Press, Oxford, 2013).

\bibitem{N10} M. E. J. Newman, \textit{Networks} (Oxford University Press, Oxford, 2010).

\bibitem{ASBS00} L. A. N. Amaral, A. Scala, M. Barth\'el\'emy and H. E. Stanley, Proc. Natl. Acad. Sci. {\bf 97}, 11149 (2000).

\bibitem{HCG16} M. O. Hase and H. L. Casa Grande, J. Stat. Mech. (2016) P043304.

\bibitem{UTGR03} P. Upham, C. Thomas, D. Gillingwater, D. Raper, J. Air Transp. Manag. {\bf 9}, 145 (2003).

\bibitem{CGCH17} H. L. Casa Grande, M. Cotacallapa and M. O. Hase, Phys. Rev. E {\bf 95}, 012321 (2017).

\bibitem{R80} L. Rayleigh, Phil. Mag. {\bf 10}, 73 (1880).

\bibitem{P05} K. Pearson, Nature {\bf 72}, 294 (1905).

\bibitem{R07} S. Redner, \textit{A Guide to First-Passage Processes} (Cambridge University Press, New York, 2007).

\bibitem{KRBN10} P. L. Krapivsky, S. Redner and E. Ben-Naim, \textit{A Kinetic View of Statistical Physics} (Cambridge University Press, New York, 2010).

\bibitem{P21} G. P\'olya, Math. Ann. {\bf 84}, 149 (1921).

\bibitem{MW65} E. W. Montroll and G. H. Weiss, J. Math. Phys. {\bf 2}, 167 (1965).

\bibitem{WS98} D. J. Watts and S. H. Strogatz, Nature {\bf 393}, 440 (1998).

\bibitem{ER59} P. Erd\H os and A. R\'enyi, Publ. Math. (Debrecen) {\bf 6}, 290 (1959).

\bibitem{NR70} H. S. Na and A. Rapoport, Math. Biosci. {\bf 6}, 313 (1970).

\bibitem{BA99} A. -L. Barab\'asi and R. Albert, Science {\bf 286}, 509 (1999).

\bibitem{OvdJ15} R. Oste and J Van der Jeugt, Electron. J. Combin. {\bf 22}, P2.8 (2015)
  
\end{thebibliography}
\end{document}